\documentclass[aoas,preprint]{imsart}

\RequirePackage[OT1]{fontenc}
\RequirePackage{amsthm,amsmath,amssymb}
\RequirePackage{natbib}
\RequirePackage[colorlinks,citecolor=blue,urlcolor=blue]{hyperref}

\usepackage{array}
\newcolumntype{C}[1]{>{\centering}p{#1}}

\usepackage{mathtools}

\usepackage{comment}
\usepackage{graphicx}
\usepackage{color}
\usepackage{float}
\usepackage{algpseudocode}
\usepackage[ruled, vlined, lined]{algorithm2e}
\usepackage{natbib}
\usepackage{macros}
\usepackage{xcolor}
\definecolor{revs}{rgb}{0, 0, 0}

\usepackage{lineno}

\startlocaldefs
\numberwithin{equation}{section}
\theoremstyle{plain}

\endlocaldefs

\usepackage{xr}
\makeatletter
\newcommand*{\addFileDependency}[1]{
  \typeout{(#1)}
  \@addtofilelist{#1}
  \IfFileExists{#1}{}{\typeout{No file #1.}}
}
\makeatother

\newcommand*{\myexternaldocument}[1]{
    \externaldocument{#1}
    \addFileDependency{#1.tex}
    \addFileDependency{#1.aux}
}
\myexternaldocument{supplement}

\begin{document}

\begin{frontmatter}
\title{Background Modeling for Double Higgs Boson Production: Density Ratios and Optimal Transport}
\runtitle{Background Modeling for Double Higgs Boson Production}

\begin{aug}
%

\author{\fnms{Tudor} \snm{Manole}
\ead[label=e1]{tmanole@andrew.cmu.edu}},
\author{\fnms{Patrick} \snm{Bryant}
\ead[label=e2]{pbryant2@andrew.cmu.edu}},
\author{\fnms{John} \snm{Alison}
\ead[label=e3]{johnalison@cmu.edu}},
\author{\fnms{Mikael} \snm{Kuusela}
\ead[label=e4]{mkuusela@andrew.cmu.edu}},
\and
\author{\fnms{Larry} \snm{Wasserman}
\ead[label=e5]{larry@stat.cmu.edu}}
\runauthor{T. Manole et al.} 

\address{Department of Statistics and Data Science and NSF AI Planning Institute for Data-Driven Discovery in Physics,\\ 
Carnegie Mellon University \\
\printead{e1,e4,e5}}

\address{Department of Physics and NSF AI Planning Institute for Data-Driven Discovery in Physics,\\
Carnegie Mellon University\\
\printead{e2,e3}\\
\phantom{E-mail:\ }}

\end{aug}

\begin{abstract}
We study the problem of data-driven background estimation, arising in the search of physics signals predicted by the Standard Model
at the Large Hadron Collider. Our work is motivated by the search for the production of pairs of Higgs
bosons decaying into four bottom quarks. A   number of other physical processes, known as background,
also share the same final state. The data arising in this problem is therefore a mixture
of unlabeled background and signal events, and the primary aim of the analysis is to determine whether the proportion of 
unlabeled signal events is nonzero.
A challenging but necessary first step is to estimate the distribution of background events.  
Past work in this area has determined regions of the space of collider events 
where signal is unlikely to appear, 
and where the background distribution is therefore identifiable. 
The background distribution can be estimated in these regions, and extrapolated into the region of primary interest
using transfer learning with a multivariate classifier.
 We build upon this existing approach in two ways. 
{\color{revs}First, we revisit this method by
developing a  customized residual neural network which is tailored to the 
structure and symmetries of collider data.}
 Second, 
we develop a new method for background estimation, based on the optimal transport problem, 
which relies on modeling assumptions distinct from earlier work. These two methods can serve as cross-checks for each other in 
particle physics analyses, due to the complementarity of their underlying assumptions. 
We compare their performance on simulated double Higgs boson data.
\end{abstract}



\end{frontmatter}

\section{Introduction} 
\label{sec:intro}
The Standard Model (SM) of particle physics is a theory describing the interactions between elementary particles---the building blocks of matter.
One key component of the SM is the presumed existence of a quantum field responsible for generating mass in certain 
elementary particles. This field is known as the Higgs field, originally theorized by \cite{higgs1964}, \cite{englert1964}. 
Excitations of the Higgs field produce particles, known as Higgs bosons, which were the subject of an 
intensive search by experimental particle physicists ever since the mid 1970s. In July 2012, two independent
experiments at the Large Hadron Collider (LHC) at CERN (the European Organization for Nuclear Research) announced the observation 
of a new particle consistent with the SM Higgs boson \citep{atlas2012, cms2012}. 
Having discovered this Higgs-like particle, current work is concerned with detailed studies of its properties,
in order to confirm or refute those predicted by the SM. 
One such property is the so-called Higgs boson \textit{self-coupling},
whereby a single excitation of the Higgs field can split into two Higgs bosons without intermediate
interactions with other particles. Observing this phenomenon would provide
compelling new information regarding the mechanism of particle mass generation.
This paper is concerned with some of the statistical challenges posed by 
its search.

The LHC is housed in a massive underground tunnel in which two counter-rotating beams of protons 
are accelerated to nearly the speed of light. When these protons collide, 
new particles are formed, and their paths within particle detectors are recorded. 
Individual collisions are referred to as \textit{events}. An event in which two Higgs bosons are generated 
is called a \textit{double Higgs (or di-Higgs) event}. The Higgs boson is a highly unstable particle; whenever it is produced, 
it decays into other particles almost immediately, making di-Higgs production impossible to observe directly. 

The Higgs boson  most commonly decays into a pair of so-called
bottom quarks ($b$-quarks). 
An event in which four $b$-quarks are observed is thus a candidate di-Higgs event, but could also 
have arisen from various other physical processes that produce four $b$-quarks. 
We say that a di-Higgs event in which the Higgs bosons decay into four $b$-quarks 
is a \textit{signal event}, while any other event tagged as having four $b$-quarks is called a \textit{background event}.
The problem of searching for double Higgs boson production reduces to testing whether the proportion of signal events is nonzero among the observed data.
As we describe in Section~\ref{sec:setup}, carrying out this test is a well-understood statistical task when 
the distributions of both background and signal events are known. 
While the di-Higgs signal distribution can be approximated to sufficient accuracy with first-principles simulation, 
simulating the background distribution suffers from large high-order corrections which are computationally 
intractable~\citep{dimicco2020}.
Instead, the background distribution must be estimated using observed data.
This is known as the problem of {\it data-driven background modeling}, which is the main subject of this paper.  

As stated, the background distribution is not a statistically identifiable quantity without further assumptions, 
due to the potential presence of an unknown proportion of signal
in the data. Any analysis strategy must therefore make some modeling assumptions to make the background estimation problem tractable. 
As we discuss below, it is standard to assume that the background distribution is related in some way to the 
distribution of certain {\it auxiliary events}, which in turn is identifiable.  
An example of useful auxiliary events is those consisting of
less than four observed $b$-quarks, since they are unlikely to be signal events, 
but are kinematically similar to the background events of interest~\citep{bryant2018}. Stated differently, the distribution of auxiliary events
is an identifiable estimand which has undergone a {\it distributional shift} relative to the non-identifiable background distribution
of interest.
If the analyst has access to a sample of auxiliary events,
its empirical distribution provides a first naive approximation of the desired background distribution.
To obtain a more precise estimate, one must correct for the distributional shift. 
 
As we discuss in Section~\ref{sec:lit},
the most widely-used method for correcting this distributional
shift  
is based on an estimate of the {\it density ratio} between the background and auxiliary
events. This method typically first estimates the density ratio
in a signal-free region of the phase space, known as the {\it Control Region}, 
and then extrapolates it to the region of primary interest, 
known as the {\it Signal Region}. 
Any deviation of this extrapolated
density ratio from unity is used
to correct the distributional shift
undergone by the auxiliary sample. 
This extrapolation can be viewed as an instance of {transfer learning}~\citep{weiss2016}. 
While a careful choice of the density ratio estimator can greatly improve the accuracy of this extrapolation, 
it clearly cannot lead to a consistent estimator  
if the  distribution in the Signal Region is unconstrained
relative to its counterpart in the Control Region.
This procedure thus places an implicit modeling
assumption on the underlying distributions, which is challenging to quantify and verify in practice. Nevertheless, variants of this procedure have been used
in each of the most recent di-Higgs searches in the four $b$-quark final state (e.g.
\cite{Aaboud_2018}, \cite{atlas2019}, \cite{Aad_2021}, \cite{CMS4b2022}, \cite{ATLAS-CONF-2022-035}). This raises the important need for  cross-checking
the  modeling assumption made by such an approach.

\subsection{Our Contributions}
This paper develops a new methodology for data-driven background modeling in di-Higgs
boson searches. Our approach is fully nonparametric, and does not involve
the extrapolation of density ratios. It hinges upon a characteristic 
modeling assumption, 
which is
complementary to that of the density ratio method. 
These two distinct methods can thus serve as cross-checks for each other in  di-Higgs   searches, an important benefit that will increase the analyst's trust in the obtained background estimates.

Our approach is based on the optimal transport problem~\citep{villani2003}
between multidimensional distributions
of collider events. Optimal transport has already proven to be a powerful tool for transfer learning in classification problems~\citep{courty2016a}, and here we propose to use it rather differently to correct distributional shifts between
estimates of the auxiliary and background distributions. 
Our method involves out-of-sample estimation of optimal transport
maps, for which we consider two different estimators. 
While the first is based on smoothing of an in-sample optimal coupling, and has previously
been proposed in the literature (cf. Section~\ref{sec:lit}), our second
estimator appears to be new, and  leverages 
some strengths of the density ratio approach.
 
The optimal transport problem requires a cost function 
on the space of collider events, for which we use a variant of  
the metric proposed by \cite{komiske2019}.
This metric is itself obtained through the 
optimal transport problem of matching clusters of 
energy deposits in collision events. 
Our approach therefore involves a nested use of~optimal~transport.

As a secondary contribution, we revisit the density ratio approach
to background estimation. In particular, we recall how 
this approach can be reduced to fitting 
a probabilistic classifier for discriminating auxiliary events from background
events, and we develop a powerful new classifier 
 tailored
to this application.  
Our classifier is a {\color{revs}customized} convolutional neural network with residual layers~\citep{he2015},
whose architecture accounts for the structure and symmetries of collider events with multiple identical final state objects.

We illustrate the empirical performance of these two methodologies
on realistic simulated collider data. We observe 
that both approaches  lead to  quantitatively similar 
background estimates, despite the complementarity 
of their underlying modeling assumptions.
In particular, this study illustrates how our 
methods can be used to cross-check each other in practice.

\subsection{Related Work}
\label{sec:lit} 

Di-Higgs boson production has been the subject of numerous recent searches by the ATLAS and CMS collaborations
at the LHC---we refer to the recent survey paper of~\cite{dimicco2020} for an overview.
The four $b$-quark final state is the most common decay channel for di-Higgs events, but suffers
from a large multijet background.
As described previously, each of the most recent searches in this final state 
 performed data-driven background estimation 
by first estimating a density ratio in a Control Region, and extrapolating it to the Signal Region. 
Certain searches, such as \cite{atlas2019}, estimate the density ratio using heuristic one-dimensional reweighting schemes, 
while others, such as \cite{CMS4b2022}, 
use off-the-shelf multivariate classifiers for this purpose. Part of our work builds upon the latter by designing a new classifier tailored to collider data.

The idea of estimating density ratios using classifiers has a long history in statistics---see for instance~\cite{fix1951}, \cite{silverman1989}, \cite{qin1998}, \cite{cheng2004}, \cite{kpotufe2017}---and   appears in a variety of applications in experimental particle physics (e.g., \cite{cranmer2015},
\cite{brehmer2020}, \cite{CMS4b2022}). 
Classification-based estimators have the practical advantage of circumventing the need for high-dimensional density estimation,
which can be particularly challenging to perform over the space of collider events.  
It has been empirically observed that modern classification algorithms, such as deep neural networks, 
have the ability to transfer well to new distributions~\citep{yosinski2014}, which further motivates
their use for density ratio estimation in our context.

Rather than density ratios, the key object of interest in our new
 methodology is  the notion of optimal transport map. 
Optimal transport theory has received a surge of recent interest
in the statistics and machine learning literature---we refer to 
\cite{panaretos2019, panaretos2019b, kolouri2017, peyre2019} for recent reviews.
Closest to our setting are applications of optimal transport to domain adaptation for classification problems; 
see for instance \cite{courty2016a}, \cite{redko2017}, \cite{rakotomamonjy2022}, and references therein. Nested optimal transport formulations, as in our work, have recently been used
for other tasks such as multilevel clustering \citep{ho2017a, ho2018, huynh2019} and  multimodal distribution alignment
\citep{lee2019}. 
Very recently, optimal transport has also been used
in high energy physics 
for calibrating stochastic simulators~\citep{pollard2022},  for purposes of
exploratory data analysis~\citep{komiske2019,komiske2020,cai2020b}, and for the
purpose of  defining a geometry on the space
of collider events~\citep{komiske2020a}. We also note that optimal transport
has implicitly been used for one-dimensional template morphing in the early work of~\cite{read1999}.

Our methodology relies on estimating optimal transport maps or couplings between
distributions of collider events. The question of out-of-sample estimation of optimal
transport maps over   Euclidean spaces has been the subject of intensive 
recent study~\citep{perrot2016, forrow2018, makkuva2019, 
nath2020,hutter2021, delara2021, deb2021, manole2021c,pooladian2021, ghosal2022, gunsilius2022}.  
While many of these works are tailored to the quadratic cost function, 
the widely-used nearest-neighbor estimator \citep{flamary2021,manole2021c} 
can naturally be defined over general metric spaces, and is used in one of our background estimators
defined in Section~\ref{sec:ot_nn_estimator}. 
 
Beyond the search of di-Higgs boson production, we emphasize that the question of data-driven background estimation
arises in a variety of problems in experimental high-energy physics, where our methodologies could also potentially
be applied. We refer to the  book \cite{behnke2013} for a pedagogical introduction to statistical aspects of the subject;
see also Appendix~1 of~\cite{lyons1986}. 
Finally, we mention some recent advances on the widely-used sPlot~\citep{barlow1987,pivk2005,borisyak2019,dembinski2022} and ABCD~\citep{Alison:1536507,ATLAS:2014aga,Choi:2019mip,Kasieczka:2020pil}
techniques for background estimation, the latter of which can be viewed as a precursor to the methods developed in this paper. 
  
\subsection{Paper Outline}
The rest of this paper is organized as follows. 
Section~\ref{sec:background} contains background about the LHC and di-Higgs boson production.
Section \ref{sec:setup} outlines the   statistical procedure used for signal searches in collider experiments at the LHC, 
and mathematically formulates the data-driven background modeling problem. In Section \ref{sec:classifier_method},
we revisit the density ratio approach to  background estimation, based on
 classifiers for discriminating auxiliary events from background events, 
and we briefly describe our new classifier architecture for this purpose.
In Section \ref{sec:ot_method}, we describe our new methodology
based on the optimal transport problem. 
We then compare these methods in a simulated di-Higgs search in Section~\ref{sec:simulation}.
We close with a discussion in Section~\ref{sec:conclusion}.  
In the Supplementary Material, Appendix~\ref*{app:hepexTranslation} contains a section-by-section summary of this manuscript
in non-technical language, Appendices~{\ref*{app:classifier_description}--\ref*{app:ot_computation}} contain
numerical details deferred from the main text, and Appendix~\ref*{app:additional_simulation_results}
contains further numerical results.

{\color{revs}
\subsection{Introduction for the High Energy Physicist}\label{sec:readersGuide}

This paper is written primarily with the statistics community in mind.
This brief section aims to bridge the gap between the language and 
formalism used by statisticians and that common in high energy physics.

We have a four-tag ($4b$) dataset consisting of background and some \textit{a priori} unknown amount of signal, parameterized by signal strength $\mu$.
The search is done in bins of a discriminating variable, 
 the output of a multivariate classifier trained to separate signal and background.
The challenge is to predict the amount of background in each of the classifier output bins.
 Section~\ref{sec:setup} introduces the three-tag ($3b$) dataset and defines the Control Region (CR) used to derive the background prediction.
The $3b$ dataset, when normalized to~the number of $4b$ events, provides a zeroth order estimate of the $4b$ background.
The main contribution of our work is in deriving corrections to the $3b$ data to better approximate the true $4b$ background.
 Section~\ref{sec:classifier_method} describes 
 a data-driven background estimation method used frequently in HEP~\citep{Aaboud_2018,CMS4b2022},
 which is a variant of the ``ABCD'' method.
We train a classifier, referred to as the ``Four-vs-Three classifier'', to obtain event weights that correct for differences between the $3b$ and $4b$ data in the CR.
The predicted background in the Signal Region (SR) is obtained by weighting $3b$ SR events by weights derived in the CR. Section~\ref{sec:ot_method} presents a novel method for data-driven background estimation.
Instead of extrapolating between the $3b$ and $4b$ samples---assuming the extrapolation is the same in the CR and SR---we propose extrapolating between the CR and SR---assuming the extrapolation is the same in the $3b$ and $4b$ samples.
We cannot use a classifier to correct kinematic differences between the samples; a classifier trained on kinematically disjoint samples would achieve perfect separation, and the corresponding weights would be undefined.
Instead, we assume that the  optimal transport map which maps events in the CR to the SR is the same for the $3b$ and $4b$ events, and we describe approaches for estimating such optimal transport
 maps with collider~data.
 
Readers less interested in the formalism may choose to skip ahead to Section~\ref{sec:simulation},
 which applies the methods introduced in this work in a simulated di-Higgs search.
A more complete description of our work aimed at the high energy physicist can be found in the Supplementary Material, Appendix~\ref*{app:hepexTranslation}.
 }

\section{Background}
\label{sec:background}

\subsection{LHC Experiments and di-Higgs Boson Production}
The LHC is the largest particle collider in the world, consisting of a 27 kilometer-long tunnel
in which two counter-rotating beams of protons are accelerated to nearly the speed of light.
These particles are primarily collided in one of four underground detectors, named ALICE, ATLAS, CMS and LHCb. 
ATLAS and CMS are general-purpose detectors used for a wide range of physics analyses, including
Higgs boson-related searches, while ALICE and LHCb focus on specific physics phenomena. 
We focus on the CMS detector in what follows, but similar descriptions can be made for the ATLAS
detector. 

\begin{figure}[!t]
\includegraphics[width=0.75\textwidth]{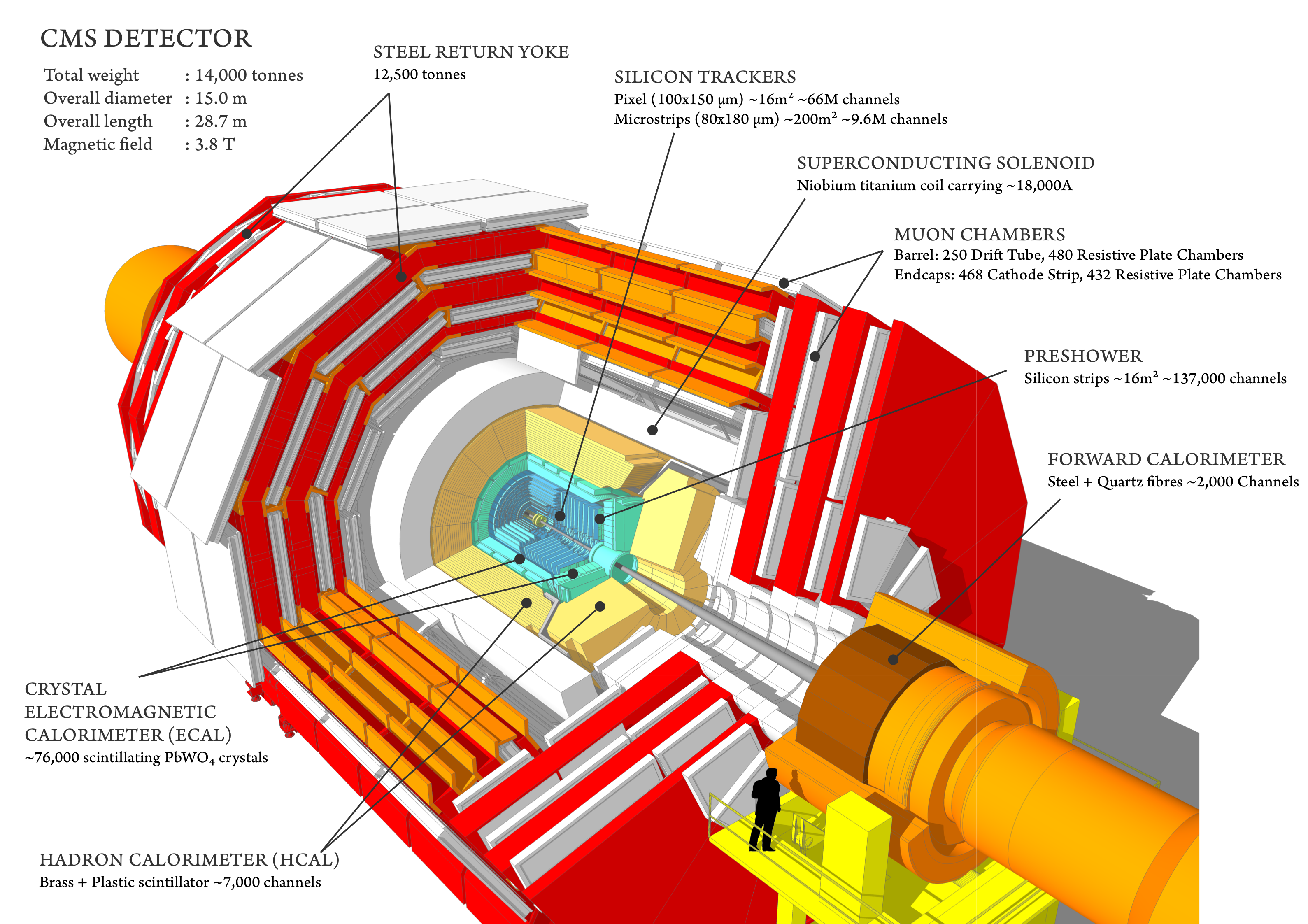}
\caption{\label{fig:cms}
Illustration of the CMS detector~\citep{sakuma2014}. Counter-rotating beams
of protons are made to collide in the center of the detector. The trajectory and mass of each particle
emanating from the collision is then recorded. }
\end{figure}

When two protons collide, their energy is converted into matter, 
in the form of new particles.
The goal of the CMS (Compact Muon Solenoid) detector is to measure the momenta,
energies and types of such particles. To measure their momenta,
CMS is built around a giant superconducting solenoid magnet, 
depicted in Figure \ref{fig:cms}, which deforms the trajectories of particles as they move from
the center of the detector outward through a silicon tracker. The extent to which the
trajectory of a charged particle is bent depends on its momentum and can hence be used to measure the momentum. After the silicon tracker, 
CMS consists of several layers of calorimeters which measure the energies of the particles.
We refer to \cite{cms2008a} for a complete description of the CMS detector. 

Proton-proton collisions give rise to highly unstable particles
which decay almost instantly into more stable particles.
The detector is only able to observe these longer-lived
particles. By measuring their energies and momenta, insight can be gained into
the physical properties of the unstable particles from which they originate.

The Higgs boson is an example of an unstable particle, which decays within
approximately $10^{-22}$ seconds. The SM predicts that 
a Higgs boson decays into a pair of bottom quarks  ($b$-quarks) $60\%$ of the time, 
and this decay channel has indeed been observed experimentally \citep{atlas2018e, cms2018b}. 
Other channels which have been observed experimentally include the decay of a Higgs boson into
pairs of
photons \citep{atlas2018, cms2018c}, W bosons \citep{atlas2018c,cms2019}, Z bosons \citep{atlas2018b,cms2018}, 
and tau leptons \citep{atlas2018d,cms2018a}. The SM further
predicts the rare possibility that two Higgs bosons can be produced simultaneously, 
and this paper is concerned with the statistical challenges arising in the search for this process, which has yet to be observed
experimentally.
If this process were to occur, the two resulting Higgs bosons would each, in turn, be
most likely to decay into two $b$-quarks, thus making four $b$-quarks the most common 
final state 
of di-Higgs boson events. We focus on this decay channel (abbreviated HH$\rightarrow{4b}$) throughout this paper.
We note that $b$-quarks form into bound states with other quarks called $b$-hadrons which are themselves unstable, and rapidly decay 
into collimated sprays of stable particles called $b$-jets, which can be efficiently identified by the CMS detector \citep{CMS:2017wtu}.

\subsection{Collider Events and the CMS Coordinate System}\label{sec:colliderEvents}
Particles measured by the CMS detector are typically
represented in spherical coordinates.  
Given a particle with momentum vector $p = (x,y,z) \in \bbR^3$, 
its azimuthal angle $\phi \in [0,2\pi)$ is defined as the angle increasing
from the positive $x$-axis to the positive $y$-axis,
while the polar angle $\theta \in [0,\pi)$ 
is increasing from the positive 
$z$-axis to the positive $y$-axis. 
The length of its projection onto the $(x,y)$ plane  is called the \emph{transverse momentum}
$p_T$. 
It is common to replace the polar angle $\theta$ by the {\it pseudorapidity} of the particle, 
given by
$\eta = -\log(\tan(\theta/2))$. 

In addition to the variables $p_T, \eta$ and $\phi$, the rest mass $m$ of each particle can be obtained
from the energy measurements made by the calorimeters in the CMS detector.
Altogether, 
a particle jet is analyzed as a single point in this coordinate system, and encoded as a four-dimensional vector $(p_T, \eta, \phi, m)$.
In our search channel, collisions lead to multiple, say $ K \geq 1$, jets measured by the detector, 
which may be encoded as the $4K$-dimensional vector $({p_T}_i, \eta_i, \phi_i, m_i:1 \leq i \leq K)$. 
We opt for an alternative notation, which will be particularly fruitful for the purpose
of defining a metric between collider events in Section \ref{sec:ot_metric}. Specifically, 
an event will henceforth be represented by the discrete measure
\begin{linenomath}\begin{equation}
\label{eq:event_measure}
g = \sum_{i=1}^K p_{T_i} \delta_{(\eta_i,\phi_i, m_i)}, 
\end{equation}\end{linenomath}
where $\delta_x$ denotes the Dirac measure placing unit mass at a point $x \in \bbR^3$. 
In particular, the representation \eqref{eq:event_measure} emphasizes the invariance of 
an event with respect 
to the ordering of its jets.
The transverse momenta $p_{T_i}$ 
may be viewed as a proxy for the energy of each jet, 
thus the total measure of $g$ denotes its total   energy, 
denoted $s_T = \sum_{i=1}^K p_{T_i}$.
The set of events with $K$ jets of interest is denoted by
\begin{linenomath}\begin{equation*}
  \calG^{(K)} = \left\{ \sum_{j=1}^K p_{T_j} \delta_{(\eta_j,\phi_j, m_j)}: p_{T_j},m_j > 0, \ \phi_j, \eta_j \in \bbR, \ 1 \leq j \leq K\right\},
\end{equation*}\end{linenomath}
where the definition of $\phi_j$ is extended from $[0,2\pi)$ to the entire real line
by $2\pi$-periodicity. 
In the context of double Higgs boson production in the four $b$-jet final state, 
the choice $K=4$ will be most frequently used, 
and in this case we simply write $\calG = \calG^{(4)}$.

Finally, we note that events are deemed invariant under the orientation of the $x$- and $z$-axes.
This fact, together with the periodicity of the angle $\phi$,
implies that two events $g = \sum_{j=1}^K p_{T_j} \delta_{(\eta_j, \phi_j, m_j)} \in \calG^{(K)}$
and $g'=\sum_{j=1}^K p_{T_j}'\delta_{(\eta_j',\phi_j',m_j')} \in \calG^{(K)}$
may be deemed equivalent if they are mirror-symmetric in $\eta,\phi$,
as well as rotationally symmetric in $\phi$,  
that is, if there exist $\Delta \in 2\pi\bbZ$
and $\iota_1, \iota_2 \in \{-1,1\}$ such that
\begin{linenomath}\begin{equation}
\label{eq:equivalence}
\sum_{j=1}^K p_{T_j} \delta_{(\iota_1 \eta_j, \Delta + \iota_2\phi_j, m_j)} = \sum_{j=1}^K p'_{T_j} \delta_{(\eta_j', \phi_j', m_j')}.
\end{equation}\end{linenomath}
Formally, we define
an equivalence relation $\simeq$ between events in $\calG^{(K)}$, such that
 $g \simeq g'$ if and only if there exist $\Delta,\iota_1,\iota_2$ for which \eqref{eq:equivalence} holds.

\section{Problem Formulation}
\label{sec:setup} 
\subsection{Overview of Signal Searches at the LHC}
\label{sec:signalSearches} 
In order to make inferences about the presence or absence of a signal process in collider data, 
event counts are commonly analyzed as binned Poisson point processes. 
While we focus
on the setting of double Higgs boson production in the four $b$-quark final state, the description
that follows is representative of a wide range of signal searches for high-energy physics experiments.

Let $\nu_0$ denote 
a $\sigma$-finite Borel measure over
the state space $\calG$ of collider events, with respect to a fixed
choice of Borel $\sigma$-algebra on $\calG$ denoted $\bbB(\calG)$.
Let $\fppp$ denote an inhomogeneous Poisson point process \citep{reiss2012} with a nonnegative intensity function $f \in L^2(\calG)$ on $\calG$,
that is, $\fppp$ is a random point measure on $\calG$ such that
\begin{enumerate}
\item $\fppp(A) \sim \text{Poisson}(\lambda(A))$, where $\lambda$ is the intensity measure induced by $f$, defined by 
$\lambda(A) = \int_A f \ddi\nu_0$ for all  $A \in \bbB(\calG)$;
\item $\fppp(A_1), \dots, \fppp(A_\ell)$ are independent for all pairwise disjoint
sets $A_1, \dots, A_\ell \in \bbB(\calG)$, for all integers $\ell \geq 1$.
\end{enumerate}
Every four $b$-jet collision event is either a \textit{signal event}, namely an
event arising from two Higgs bosons, or a \textit{background event}, arising from some other physical process.
Letting $\mu\geq 0$ denote the rate of signal events, we write the intensity measure $\lambda$ as 
\begin{linenomath}
\begin{equation*}
\lambda(\cdot) = \binten_4(\cdot) + \mu \sinten(\cdot), \end{equation*}\end{linenomath}  
where $\binten_4$ and $\sigma$, respectively, denote nonnegative background and signal intensity measures.
$\sigma$ is typically normalized
such that the value $\mu=1$ corresponds to the theoretical prediction of 
the signal rate.
The measures $\binten_4$ and $\sigma$ typically depend on nuisance parameters 
related to the calibration of the detector, 
the uncertain parameters of certain physical processes, such as the parton distribution functions of the proton 
\citep{placakyte2011}, and so on. We suppress the dependence on 
such nuisance parameters for ease of exposition.
The parameter $\mu$ is of primary interest, since non-zero values of $\mu$ indicate
the existence of signal events. A search for the signal process therefore reduces to
testing the following hypotheses on the basis of observations from the Poisson point process $\fppp$:
\begin{linenomath}\begin{equation}
\label{eq:hypotheses}
H_0: \mu = 0 \quad \text{ vs. } \quad H_1: \mu > 0.
\end{equation}\end{linenomath}

Given a sequence $G_{1}, G_{2}, \dots$ of observed events, we may write $\fppp = \sum_{i=1}^{\Nfppp} \delta_{G_i}$,
where $\Nfppp \sim \text{Poisson}(\lambda(\calG))$
is independent of the observations, which satisfy
\begin{linenomath}\begin{equation}
\label{eq:huber}
G_1, G_2, \dots \overset{\text{iid}}{\sim}  \lambda / \lambda(\calG) = \epsilon S + (1-\epsilon)  P_4.
\end{equation}\end{linenomath}
Here, $S = \sigma/\sigma(\calG)$ and $P_4 = \binten_4/\binten_4(\calG)$ denote the respective  signal and background distributions, and 
$\epsilon = \mu \sigma(\calG)/\lambda(\calG)$ the  proportion of signal events.

The Poisson point process $\fppp$ is often binned in practice. 
Let $\xi:\calG \to \calA \subseteq \bbR$ denote a dimensionality reduction map,
to be discussed below, 
which will be used to bin the point process using univariate bins.
Let $\{I_j\}_{j=1}^J$ denote a collection of bins forming a partition of $\calA$, and define
the event counts
\begin{linenomath}\begin{equation}
\label{eq:counts}
D_j = \fppp\big(\xi^{-1}(I_j)\big) = \big|\{1 \leq i \leq \Nfppp:  \xi(G_i) \in I_j \}\big|, \quad j=1, \dots, J.
\end{equation}\end{linenomath}
The definition of $\fppp$ implies that the random variables $D_j$ 
are independent and satisfy
\begin{linenomath}\begin{equation}
\label{eq:poisson_assm}
D_j \sim \text{Poisson}\big(B_j  + \mu S_j\big), \quad j=1, \dots, J,
\end{equation}\end{linenomath}
where $B_j = \binten_4 (\xi^{-1}(I_j))$ and $S_j = \sinten (\xi^{-1}(I_j))$.

The likelihood ratio test with respect to the joint distribution of $D_1, \dots, D_J$
is typically used to test the hypotheses \eqref{eq:hypotheses} \citep{atlas2011}. 
The binned likelihood function for the parameter $\mu$ is   given by
\begin{linenomath}\begin{equation}
\label{eq:likelihood}
L(\mu) = \prod_{j=1}^J \frac{\big(B_j + \mu S_j\big)^{D_j}}{D_j!} 
e^{-\big(B_j+\mu S_j\big)}.
\end{equation}\end{linenomath}
Di-Higgs events are rare in comparison to background events, thus the signal-to-background
ratio is low. At the time of writing,  values of 
$\nf$ which are typically observed at the LHC may be too small  
for any test to 
have power in rejecting the null hypothesis in \eqref{eq:hypotheses} at desired significance levels~\citep{dimicco2020}. 
Analyses which fail to reject $H_0$ instead culminate in an upper confidence bound on $\mu$, also known as an upper limit 
\citep{atlas2011}.

The power of the likelihood ratio test for \eqref{eq:hypotheses} may be increased by choosing a function~$\xi$
which maximizes the separation between background and signal event counts across the $J$ bins. 
Informally, the optimal such choice of $\xi$ is  given by  
\begin{linenomath}\begin{equation}
\label{eq:signal_classifier_setup}
\xi(g) = \bbP(G \text{ is a Signal Event} | G = g),
\end{equation}\end{linenomath}
which may be estimated using a multivariate classifier, such as a neural network or boosted decision trees, for discriminating background events from signal events. 

The signal intensity measure $\sigma$ is theoretically predicted by the SM, 
and can be approximated well using Monte Carlo event generators~\citep{dimicco2020}. 
The background intensity $\binten_4$ is, however, 
intractable due to the strongly interacting nature of quantum chromodynamics (QCD) in which
events with the four $b$-quark final state can be produced
via an enormous number of relevant and complex pathways.
The intensity measure $\binten_4$, or its binned analogue $(B_j)_{j=1}^J$,  must therefore be estimated from the collider data itself, which 
we refer to as \textit{data-driven background modeling}. This problem is the primary focus of this paper.

\subsection{Setup for Data-Driven Background Modeling}
\label{sec:setup_background}

The aim of this paper is to develop data-driven estimators of the background intensity measure $\binten_4$. 
The primary challenge is 
the fact that the sample $G_1, \dots, G_{\nf}$  
is contaminated with an unknown proportion $\epsilon$ of signal events. 
The background estimation problem is thus  statistically unidentifiable as stated,
and it will be necessary to impose further modeling assumptions.

In order to formulate these assumptions and our resulting background modeling
methods, 
we assume that the analyst has access to a second Poisson Point Process $T=\sum_{i=1}^{\Ntppp} \delta_{H_i}$
consisting of auxiliary events which were tagged by the CMS detector as
having four jets, of which exactly three are $b$-jets. We refer to such
events as ``$3b$ events'', as opposed to ``$4b$ events'' which were identified as having four $b$-jets\footnote{$3b$ events were used for background estimation in the HH$\rightarrow 4b$ channel
in the recent analysis of~\cite{CMS4b2022}. ``$2b$ events'' consisting of two, rather
then three, $b$-tagged jets
have  been used in other recent analyses (e.g.~\cite{atlas2019,Aad_2021}), and our description also applies to such events with only formal changes.}.
We stress that the terms $3b$ and $4b$ do not refer to the true number of $b$-quarks arising from the collision, 
rather the number of $b$-jets identified by the detector.
As we   discuss in Section~\ref{sec:simulation}, the majority of $3b$ events~in~fact~arise from the hadronization of two $b$-quarks
and two charm or light quarks,   
while a small proportion arise 
from four $b$-quarks\footnote{As a result, the expected
rate of production of $3b$ events $\bbE[N]$ is typically higher
than that of $4b$ events $\bbE[M]$ by an order
of magnitude; cf. Section~\ref{sec:simulation}}. For the purpose of a discovery analysis, 
the $3b$ sample $H_1, \dots, H_{\Ntppp}$ can therefore be treated
as having a negligible proportion of signal events~\citep{bryant2018,CMS4b2022}.
We treat this proportion as zero for sake of exposition.
We   henceforth denote the intensity measure of the point  process $T$ by $\beta_3$, and we denote by $P_3 = \binten_3/\binten_3(\calG)$ the corresponding probability distribution of the  observations~$H_1, H_2, \dots$.

The kinematics of $3b$ events are   
similar, but not equal, to those of $4b$ background events~\citep{CMS4b2022}. 
Unlike $\binten_4$, however, the intensity measure $\binten_3$ 
is an identifiable estimand  due
to the lack of signal events in the point process $T$. 
Any consistent estimator $\hat \binten_3$ of $\binten_3$ can
be used to provide a zeroth-order approximation of $\beta_4$ (up to a correction for normalization).
This approximation is, however, insufficiently 
accurate to be used as a final estimate of $\binten_4$
and our goal is to develop statistical methods for  correcting this naive background estimate. 

Recall  that the four $b$-jets of any signal event $g 
\in \calG$ are naturally paired, with each pair
arising from a Higgs boson. The true pairing of the jets is unknown
to the detector; however, it may be approximated, for instance using an algorithm due to \cite{bryant2018}. 
We use the same pairing algorithm in our work.
Given as input an event $g$, this deterministic algorithm outputs one among the
three distinct unordered pairs of measures $\{g^1,g^2\}\subseteq \calG^{(2)}$
which satisfy $g = g^1+g^2$.  
We refer to $g^1$ and $g^2$ as {\it dijets}.
When $g$ is a signal event, we expect that each dijet
arose from a decay of a Higgs boson, whereas when $g$ is a background event, 
we expect that at least one of the two dijets arose from the decay of a different particle.

The Higgs boson is known to have mass $m_H$ approximately equal to 125 GeV
\citep{atlas2012, cms2012}. It follows that
the two dijets should approximately satisfy 
$m(g^1) \approx m(g^2) \approx m_H$,
where $m(a)$ denotes the invariant mass\footnote{If $E$ denotes
the sum of the energies of the constituent jets of  $a$, 
and $p$ denotes the magnitude of the sum of their momentum vectors, then
the invariant mass of $e$ is defined by 
$m(a) = \sqrt{E^2 - p^2}$~\citep{hagedorn1964}.  } 
 of any $a \in \calG^{(K)}$, $K \geq 1$.  
Large deviations
of the dijet invariant masses from 125 GeV indicate that $g$ is not a signal event. This provides
a heuristic for determining events among $G_1, \dots, G_M$ 
which are unlikely to be signal events. 
To elaborate, we form subsets $\calGcr, \calGsr \subseteq \calG$  
such that $\calGcr \cap \calGsr = \emptyset$, where  
$\calGsr$ is called the \textit{Signal Region}, containing 
events with dijet masses near $m_H$, and $\calGcr$ is called the \textit{Control Region}, 
containing all other events which will be used in the analysis. We follow 
\cite{bryant2018} and employ the following specific definitions of $\calGcr$ and $\calGsr$:
\begin{linenomath}\begin{align}
\label{eq:signal_region_defn} 
\calGsr &= \left\{ g \in \calG: 
\sqrt{\left(1 - \frac{ m_H}{m(g^1)}\right)^2 + 
      \left(1-\frac{m_H}{ m(g^2)}\right)^2}
\leq \kappa_s\right\},\\ 
\label{eq:control_region_defn}
\calGcr &= \left\{ g \in \calG:  \sqrt{\Big(m(g^1) - \sigma_c m_H  \Big)^2 + 
                                                 \Big(m(g^2) - \sigma_C m_H \Big)^2}
\leq {\color{revs}r_c}\right\}  \setminus \calGsr,
\end{align}\end{linenomath}
for some constants $\sigma_c, {\color{revs}r_c}, \kappa_s > 0$.  
\begin{figure}[t]
\includegraphics[width=0.4\textwidth]{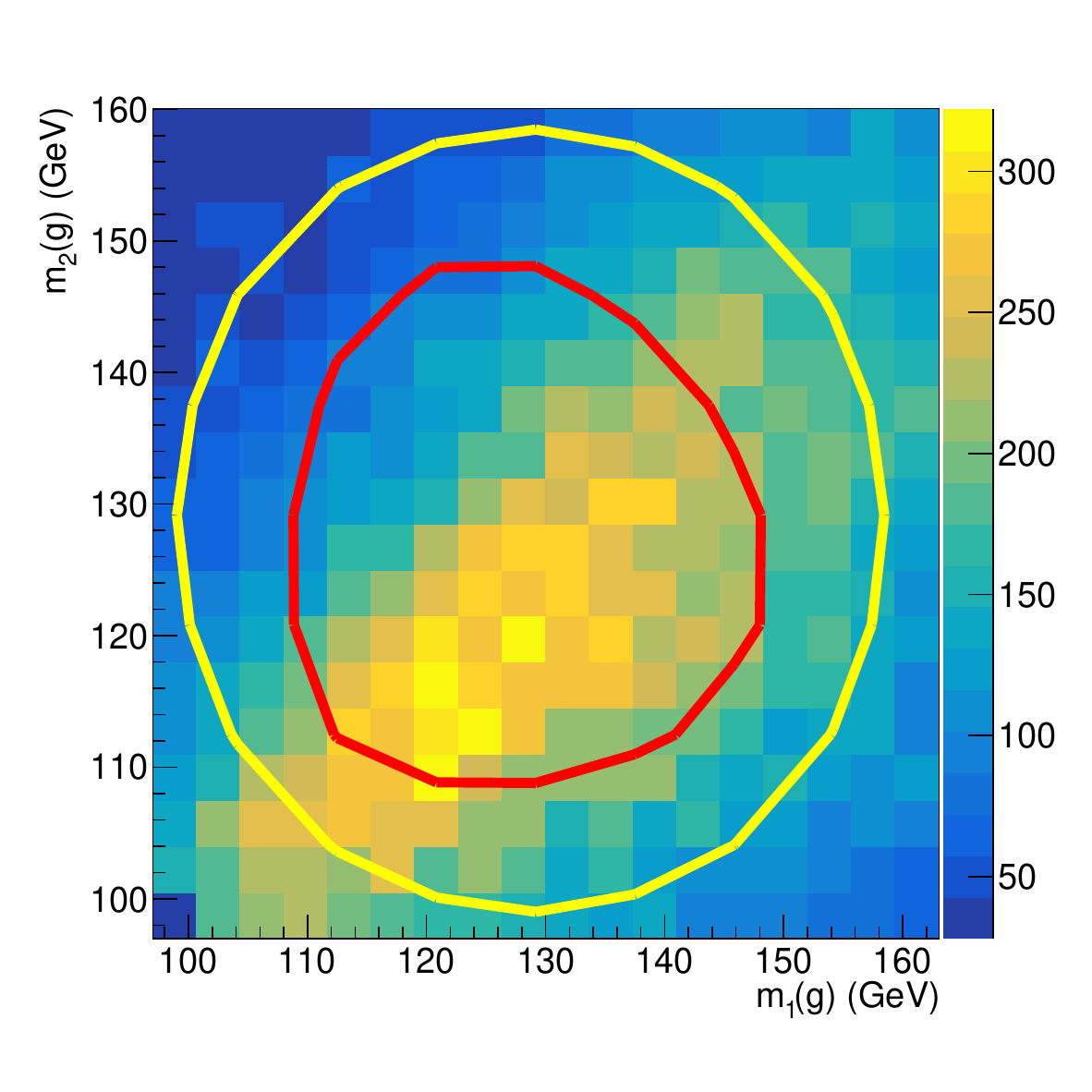}
\caption{\label{fig:regions} Illustration of the Control and Signal Regions.
The two-dimensional histogram represents simulated 4b collider events described
in Section~\ref{sec:simulation}, plotted in terms of their dijet invariant masses. We emphasize
that this is a low-dimensional representation; the events considered in this work are 16-dimensional. 
The red line indicates the boundary of the Signal Region, while the annulus bounded by the yellow and red lines represents
the Control Region. The constants $\sigma_c, {\color{revs}r_c}$ and $\kappa_s$
used in this figure are stated in Section~\ref{sec:simulation}. 
}
\end{figure} 
These regions are illustrated in Figure \ref{fig:regions}.
We similarly partition the Poisson intensity measures $\beta_3, \beta_4$, by defining
for all $A \in \bbB(\calG)$, 
\begin{linenomath}
\begin{equation*}
\beta_j^c(A) = \beta_j(A \cap \calGcr), \quad 
\beta_j^s(A) = \beta_j(A \cap \calGsr), \quad  j=3,4.
\end{equation*}
\end{linenomath}
  These four measures are illustrated in Figure~\ref{fig:all_methods}.
Furthermore, we assume for ease of exposition that these measures are absolutely
continuous with respect to the dominating measure $\nu_0$, 
and we let  $b_j^a = d\beta_j^a/d\nu_0$ for all $j =3,4$ and $a=c,s$. 
  
Recall that the collider events associated
with the intensity measures $\binten_3^c$ and $\binten_3^s$ are signal-free by construction, 
and those from $\binten_4^c$ are also signal-free under the assumption that 
negligibly few signal events will fall outside of $\calGsr$. These three intensity measures can
therefore be estimated directly by means of their empirical intensity functions. 
We have thus reduced the background modeling problem to that of estimating $\binten_4^s$,
given estimates of $\binten_3^c$, $\binten_3^s$ and $\binten_4^c$.
 
To this end, we will partition the samples 
into the sets 
\begin{alignat*}{4}
&\{G_1^s, \dots, G_{\nfs}^s\} &&:= \{G_1, \dots, G_{\nf}\} \cap \calGsr,
     \qquad &&\{H_1^s, \dots, H_{\nts}^s\} &&:= \{H_1, \dots, H_{\nt}\} \cap \calGsr, \\
& \{G_1^c, \dots, G_{\nfc}^c\}&&:= \{G_1, \dots, G_{\nf}\} \cap \calGcr,   
             \qquad && \{H_1^c, \dots, H_{\ntc}^c\}&&:= \{H_1, \dots, H_{\nt}\} \cap \calGcr,  
\end{alignat*}
where $\nf = \nfc+\nfs$ 
and $\nt = \ntc+\nts$. 
Furthermore, let 
\begin{linenomath}\begin{equation*}\binten^c_{3,\ntc} = \tppp|_{\calGcr} =  \sum_{i=1}^{\ntc} \delta_{H_i^c},\quad 
  \binten^s_{3,\nts} = \tppp|_{\calGsr} =  \sum_{i=1}^{\nts} \delta_{H_i^s},\quad 
  \binten^c_{4,\nfc} = \fppp|_{\calGcr} =  \sum_{i=1}^{\nfc} \delta_{G_i^c}\end{equation*}\end{linenomath}  
denote the  empirical estimators
of the measures $\binten_3^c, \binten_3^s, \binten_4^c$, illustrated in the background of Figure~\ref{fig:all_methods}.
As previously noted, the measure $\binten_3^s$ provides a zeroth-order 
approximation of $\binten_4^s$ (after a normalization correction), 
thus a naive first estimate of $\binten_4^s$ is given by $\binten^s_{3,\nts}$.   
As we shall see in the simulation study of Section \ref{sec:simulation}, this approximation is insufficiently
accurate to be used as a final estimator. Our methodologies improve upon
it by modeling the discrepancy between the $3b$ and $4b$ distributions in the Control Region 
via $\binten_{4,\nfc}^c, \binten_{3, \ntc}^c$, and then using that information in the Signal Region to improve the accuracy of $\binten^s_{3,\nts}$ as an estimator of~$\binten_4^s$.

\begin{figure}[t]
\includegraphics[width=0.4\textwidth]{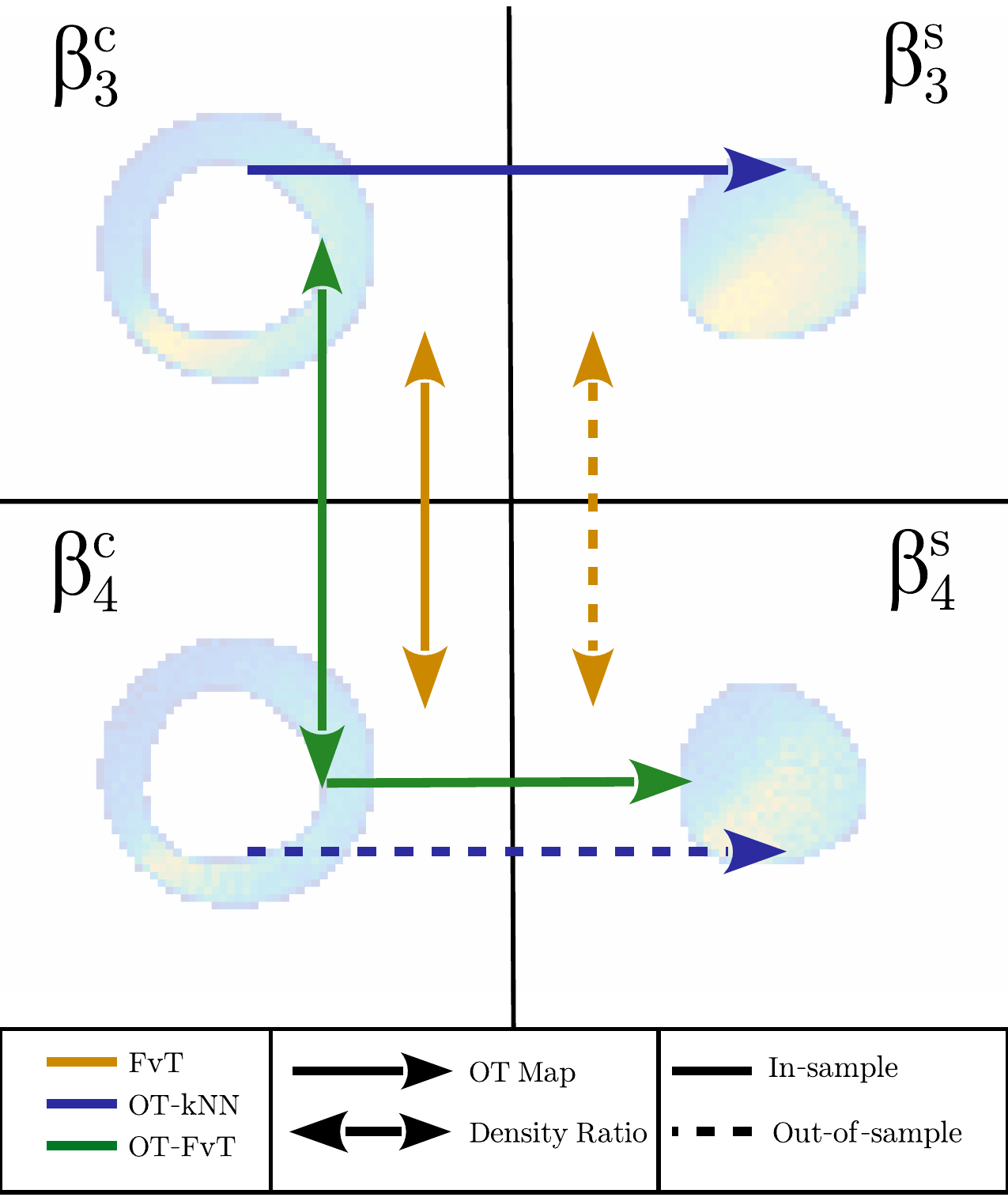}
\caption{\label{fig:all_methods} 
{\color{revs}(Color online.)}
Illustration of the four Poisson intensity measures 
$\binten_3^c, \binten_3^s, \binten_4^c, \binten_4^s$, among which only
the latter is nontrivial to estimate, and summary of the three methods developed in this paper for estimating
 $\binten_4^s$. 
The method FvT (Four vs. Three) estimates the ratio of the two densities in the Control Region using a classifier, 
and then extrapolates it into the Signal Region
using out-of-sample evaluations of the classifier.
The OT-$k$NN (Optimal Transport--$k$ Nearest Neighbors) 
method produces an estimator $\hat T$ of the optimal transport map $T$ between the $3b$ Control
and Signal Region distributions, and evaluates this estimator out-of-sample on an estimator of the 
4b Control Region distribution. The out-of-sample evaluation of $\hat T$ is performed using nearest-neighbor extrapolation.
The OT-FvT (Optimal Transport--Four vs. Three) method combines these ideas: 
first, it uses the classifier to produce an estimator of $\binten_4^c$ with the same support as
$\binten_{3,\ntc}^c$, and second, it pushes forward this estimator through $\hat T$, 
thereby avoiding out-of-sample 
evaluations of both the classifier and optimal transport map. The background of the figure
consists of bivariate histograms of simulated $3b$ and $4b$ samples in the Control and Signal Regions, 
plotted in terms of their dijet invariant masses, as in Figure~\ref{fig:regions}. }
\end{figure} 
Once we are able to derive an estimator $\hat\binten_{4}^s$ of $\binten_4^s$, based on 
the signal-free observations $G_1^c, \dots, G_{\nfc}^c, $ $H_1^s, \dots, H_{\nts}$, $H_1^c, \dots, H_{\ntc}^c$,
we may define the fitted histogram $\hat B_j = \hat\binten_4^s(\xi^{-1}(I_j))$, $j=1, \dots, J$. One may then
test the hypotheses \eqref{eq:hypotheses} using the likelihood ratio test, based on the following
modification of the likelihood function in equation~\eqref{eq:likelihood},
\begin{linenomath}\begin{equation}
\label{eq:modified_likelihood}
\widetilde L(\mu) = \prod_{J=1}^J \frac{\left( \hat B_j + \mu S_j\right)^{D_j^s}}{D_j^s!} e^{-\left(\hat B_j + \mu S_j\right)},
\quad 
\text{where } D_j^s = \left|\left\{1 \leq i \leq m_s: \xi(G_i^s) \in I_j\right\}\right|.
\end{equation}\end{linenomath}
Here, $\widetilde L$ can be viewed as a restriction of the likelihood $L$ to the Signal Region.
Notice that $\hat B_j$ is independent of $D_k^s$, for any $j,k$. 
In practice, it is also necessary to incorporate statistical and systematic uncertainties pertaining to the estimator $\hat B_j$
into the hypothesis testing procedure \citep{atlas2011}. Since formal uncertainty quantification for background
modeling is beyond the scope of this work, we omit further details, and provide further discussion of this point in Section~\ref{sec:conclusion}.
 
 The primary difficulty remaining in the testing problem~\eqref{eq:hypotheses}
is that of deriving estimators of the background intensity measure 
$\binten_4^s$. In what follows, we describe two classes of estimators for $\binten_4^s$:
one based on density ratio estimation (Section~\ref{sec:classifier_method}),
and the second based on  optimal
transport   (Section~\ref{sec:ot_method}). 
The former is the most common approach to di-Higgs background modeling, 
and will later be referred to as the FvT method. 
The latter is new, and we will discuss two different instances
of this approach, which
will later
be referred to as the OT-$k$NN and OT-FvT methods.
These three distinct estimators are summarized in Figure~\ref{fig:all_methods}.

\section{Background Modeling via Density Ratio Extrapolation}
\label{sec:classifier_method} 
The discrepancy between $3b$ and $4b$ background distributions may be directly quantified
in the Control Region, where no signal events are present.
Under 
a suitable modeling assumption, 
this discrepancy may be extrapolated into the Signal Region 
to produce a correction of  the $3b$ signal region intensity measure $\binten_3^s$, leading to an estimate
of $\binten_4^s$. 
This general strategy forms the basis of most background modeling
methodologies used in recent di-Higgs searches, as discussed in Section~\ref{sec:intro}.  
The aim of this section is to recall
how this approach may be carried out using a classifier
for discriminating $3b$ and $4b$ events. We then {\color{revs}propose a  
classifier} specifically tailored to this type of collider data, 
which will be used in our numerical studies.

Let $E$ denote a random collider event, arising from either
the $3b$ or $4b$ distributions,
and define the latent binary random variable $Z$ indicating the component membership
of $E$. More specifically,  
let $Z$ be a Bernoulli random variable with
success probability $\bbP(Z=1) = \binten_4(\calG)/(\binten_4(\calG)+\binten_3(\calG))$, and let $E$ be
 generated according to the mixture model
$$E|Z=0 \sim P_3, \quad
E|Z=1 \sim P_4.$$ 
Setting 
$\psi(g) = \bbP(Z=1|E=g)$ for all $g \in \calG$,  
it follows from Bayes' Rule that 
\begin{linenomath}\begin{equation}
\label{eq:ratio_classifier_density}
\frac{b_4^c(g)}{b_3^c(g)} = \frac{\psi(g)}{1-\psi(g)},\quad g \in \calGcr,
\end{equation}\end{linenomath}
where we recall that $b_j^c$ denotes the intensity function associated to $\binten_j^c$, $j=3,4$. 
Therefore,
\begin{linenomath}\begin{equation}
\label{eq:ratio_classifier}
\binten_4^c(A) = \int_A \frac{\psi(g)}{1-\psi(g)} \ddi \binten_3^c(g),\quad A \in \bbB(\calGcr).
\end{equation}\end{linenomath}
Equations (\ref{eq:ratio_classifier_density}--\ref{eq:ratio_classifier}) are a reformulation for our context
of the well-known fact that, up to normalization, 
a likelihood ratio may be expressed as an odds 
ratio~\citep{silverman1989}. 
Estimating the ratio of $3b$ to $4b$ intensity functions
in the Control Region thus reduces
to the classification problem of estimating $\psi$, say by a classifier~$\hat \psi$.
This observation has the practical advantage of circumventing 
the need of performing high-dimensional
density estimation. 
Assuming that the estimator~$\hat \psi$ can be evaluated in
the Signal Region $\calGsr$, disjoint from its training 
region $\calGcr$, we may postulate that the measure  
\begin{linenomath}\begin{equation}
\label{eq:ansatz}
A \in \bbB(\calGsr) \longmapsto \int_A \frac{\hat \psi(g)}{1-\hat \psi(g)} \ddi\binten_3^s(g)
\end{equation}\end{linenomath}
provides a reasonable approximation of $\binten_4^s$.
The quality of such an approximation is   driven by the ability of the classifier $\hat \psi$ 
to generalize between regions of the phase space.
To formalize this, we will assume for simplicity that
$\hat\psi$ is an empirical risk minimizer taking values in a class $\{\psi_\alpha:\calG \to [0,1]: \alpha \in \Omega\}$, 
for some parameter space $\Omega \subseteq \bbR^d$, $d \geq 1$.
That is, we assume
\begin{linenomath}\begin{equation*}\hpsi = \psi_{\halpha}, ~~ \text{where} ~~
\halpha = \argmin_{\alpha \in \Omega} \left\{
	\frac 1 {\ntc} \sum_{i=1}^{\ntc} \calL\big(\psi_\alpha(H_i^c), 0\big) + 
	\frac 1 {\nfc} \sum_{j=1}^{\nfc} \calL\big(\psi_\alpha(G_j^c), 1\big)\right\},\end{equation*}\end{linenomath}
for some loss function $\calL:[0,1]\times [0,1] \to \bbR$. 
We then make the following assumption:
\begin{assumption}
\label{assm:classifier}
The conditional probability $\psi$   satisfies the following conditions:
\begin{itemize}
\item[(i)] (Correct Specification) There exists $\alpha^* \in \Omega$ such that $\psi = \psi_{\alpha^*}$.
\item[(ii)](Generalization) We have 
\begin{linenomath}\begin{equation*}\alpha^* = \argmin \limits_{\alpha \in\Omega} \bbE\big[\calL(\psi_\alpha(G), Z) | G \in \calGcr\big].\end{equation*}\end{linenomath}
\end{itemize}
\end{assumption}
Assumption \ref{assm:classifier} implies that a classifier trained solely in the Control Region can consistently
estimate the full conditional probability $\psi(g)$, for events $g \in \calG$ lying in both the Control and Signal Regions.  
Such an assumption guarantees the ability of the classifier $\hpsi$ to generalize from the Control Region, 
making the ansatz \eqref{eq:ansatz} justified. 
A natural estimator for $\beta_4^s$ is then
obtained by replacing $\binten_3^s$ in equation~\eqref{eq:ansatz} by its empirical counterpart
$\binten_{3,\nts}^s$. Doing so leads to the estimator
\begin{linenomath}\begin{equation}\label{eq:FvTestimator}\hat \binten^s_4 
 = \sum_{i=1}^{\nts} \frac{\hat \psi(H_i^s)}{1-\hat \psi(H_i^s)} \delta_{H_i^s}.\end{equation}\end{linenomath}
$\hat \binten^s_4$ is called the \fvt estimator, and we refer to  
$\hat\psi$   as the FvT (Four vs. Three) classifier.

The validity of Assumption~\ref{assm:classifier} relies crucially upon the choice of the
function class $\{\psi_\alpha\}$,~or equivalently the choice of the classifier
$\hat\psi$. 
Indeed, off-the-shelf classifiers may lack the generalization ability to 
satisfy Assumption~\ref{assm:classifier}(ii). 
A secondary contribution of our work is to propose a 
classifier 
 specifically tailored to 
four-jet collider events, which we now introduce.

 \vspace{0.05in}
\noindent {\bf The FvT Classifier.} 
Our aim is to design a classifier $\hat \psi$ over $\calG$, which 
\begin{enumerate}
\item[(a)] is invariant to the ordering of the constituent jets in an input event $g$;
\item[(b)] is invariant with respect to the equivalence relation $\simeq$ defined in \eqref{eq:equivalence};
\item[(c)] incorporates the dijet substructure of an event $g = g^1+g^2$. 
\end{enumerate}
In Appendix~\ref*{app:classifier_description}, we describe how these properties can be satisfied using
a customized convolutional neural network architecture with residual layers, or ResNet \citep{he2015}.
We refer to the resulting classifier as the FvT classifier, and implement the
 FvT method with this choice throughout our numerical studies in Section~\ref{sec:simulation}. 
 Beyond 
its use for background modeling, 
we also employ this  classifier    for the final dimensionality reduction map $\xi$
in equation \eqref{eq:signal_classifier_setup}. 
Choosing these two classifiers to have the same architecture is important 
in practice, since a classifier capable of learning the relevant features for signal extraction should also be capable of learning and then correcting those same features in the background model.

\section{Background Modeling via Optimal Transport}
\label{sec:ot_method}
The methodology described in the previous section hinged 
  upon the ability of the classifier $\hat\psi$ to accurately
extrapolate from
the Control Region to the Signal Region, implying that the 
$3b$ and $4b$ intensity functions in the latter region are constrained
by their values in the former region.
The validity of this assumption is difficult to verify
in practice due to the blinding of the $4b$ signal region 
which motivates us to develop a distinct approach with a complementary modeling
assumption. In this section, rather than extrapolating the discrepancy between the $3b$ and $4b$ 
intensity functions, we will  
extrapolate the discrepancy between the Control and Signal Region intensity functions, as illustrated
in Figure \ref{fig:all_methods}.

We cannot use a density ratio to quantify the discrepancy between 
the intensity functions in the Control and Signal Regions, because these regions are disjoint. 
We will instead use the notion of a {\it transport map}, which will be defined below. 
In order to employ transport maps, it will be convenient  to normalize all intensity functions throughout this section. 
That is, we will define an estimator for  $\binten_4^s$ by 
separately estimating the probability measure $P_4^s = \binten_4^s/\binten_4^s(\calGsr)$ and the normalization $\binten_4^s(\calGsr)$.  
More generally, we denote by
\begin{linenomath}\begin{equation*}P_j^c = \binten_j^c / \binten_j^c(\calGcr), \quad
  P_j^s = \binten_j^s / \binten_j^s(\calGsr), \quad j=3,4,\end{equation*}\end{linenomath}
 the four population-level probability measures, with corresponding empirical measures
\begin{linenomath}\begin{equation*}P_{3,n_a}^a = \frac 1 {n_a} \sum_{i=1}^{n_a} \delta_{H_i^a},\quad 
  P_{4,m_a}^a = \frac 1 {m_a} \sum_{i=1}^{m_a} \delta_{G_i^a}, \quad a \in \{c,s\}.\end{equation*}\end{linenomath}
  
A transport map~\citep{villani2003} between $P_3^c$ and $P_3^s$ is any Borel-measurable function $T: \calGcr \to \calGsr$
such that whenever $H \sim P_3^c$, we have $T(H) \sim P_3^s$. When   this condition holds,  
we write $P_3^s = T_\# P_3^c$, and we say $T$ {\it pushes $P_3^c$ forward} onto $P_3^s$, or that $P_3^s$ is the {\it pushforward}
of $P_3^c$ under~$T$. 
Equivalently, this condition holds if and only if
\begin{linenomath}
\begin{equation*}
P_3^s(A) = T_\# P_3^c(A) = P_3^c(T^{-1}(A)),\quad \text{for all } A \in \bbB(\calGsr).
\end{equation*}\end{linenomath}
We propose to perform background estimation under the following informal modeling assumption, which will be stated more formally in the sequel. 
\refstepcounter{assump}\label{2}
\begin{assump}  
\label{assm:ot_pre}
There exists a map $T_0:\calGcr \to \calGsr$ such that
\begin{linenomath}\begin{equation}
\label{eq:pushforward_assm2'}
{T_0}_\# P_3^c = P_3^s, \quad \text{and}\quad {T_0}_\# P_4^c = P_4^s.
\end{equation}\end{linenomath}
\end{assump}  
Assumption~\ref{assm:ot_pre} requires
the  $3b$ and $4b$ distributions to be sufficiently similar
for there to exist a shared map $T_0$ which pushes forward
their restrictions to the Control Region  into their counterparts in the Signal Region. 
If such a map $T_0$ were available, it would suggest the following procedure for 
estimating $P_4^s$:
\begin{enumerate}
\item [(a)] Fit an estimator $\hat T$ of $T_0$ based only on the $3b$ observations;
\item [(b)] Given any estimator $\hat P_{4,\nfc}^c$ of $P_{4}^c$, 
use the pushforward $\hat T_\# \hat P_{4,\nfc}^c$ as an estimator of $P_4^s$. 
\end{enumerate}
For this approach to be practical, we must specify an explicit candidate  $T_0$ satisfying Assumption~\ref{assm:ot_pre}. 
We propose to choose $T_0$ such that its
movement of the probability mass from $P_3^c$ into that of $P_3^s$ is minimal. This leads us to consider the classical optimal transport problem, 
which we now describe. 

 \subsection{The Optimal Transport Problem}\label{sec:OTProblem}
Assume a metric $W$ 
on the space $\calG$ is given; we provide a candidate for such a metric in Section~\ref{sec:ot_metric}. 
For any transport map $T$ pushing $P_3^c$ forward onto $P_3^s$, 
we refer to $W(h,T(h))$ as the cost of moving an event 
$h \in \calGcr$ to an event $T(h) \in \calGsr$.  
The optimal transport problem seeks to find the choice of $T$ which minimizes the expected 
cost of transporting $P_3^c$ onto $P_3^s$, which amounts to solving the following optimization problem
\begin{linenomath}\begin{equation}
\label{eq:monge}
\argmin_{T: \calGcr \to \calGsr} \int_{\calGcr} W(h, T(h)) \ddi P_3^c(h), \quad \text{s.t. }  T_\# P_3^c = P_3^s.
\end{equation}\end{linenomath}
Equation \eqref{eq:monge} is known as the Monge problem \citep{monge1781}. 
When a solution $T_0$ to the Monge problem exists, it is said to be 
an \textit{optimal transport map}. We postulate that, when it exists, the optimal
transport map from $P_3^c$ to $P_3^s$ is a sensible
candidate for the map $T_0$ appearing in the statement of Assumption~\ref{assm:ot_pre}.

A shortcoming of this choice is the requirement that there exist a solution to the optimization problem~\eqref{eq:monge}. 
It is well-known that the Monge problem over Euclidean space admits a unique solution for  absolutely
continuous distributions, when the cost function is the squared
Euclidean norm~\citep{knott1984,brenier1991}. 
While sufficient conditions for the solvability of the Monge problem in more general spaces
are given by~\citeauthor{villani2008} (\citeyear{villani2008}, Chapter~9), we do not know whether they
are satisfied by the metric space $(\calG, W)$ under consideration. 
Furthermore, the Monge problem may not even  be feasible between distributions which are not absolutely continuous, 
which precludes the possibility of estimating $T_0$ using the optimal transport map
between the empirical measures of $P_3^c$ and $P_3^s$. 

Motivated by these considerations, we introduce a classical relaxation of the Monge problem, 
known as the Kantorovich optimal transport problem \citep{kantorovich1942, kantorovich1948}.
Let $\Pi(P_3^c, P_3^s)$ denote the set of all joint Borel distributions $\pi$ over $\calGcr\times \calGsr$ whose
marginals are, respectively, $P_c^3$ and $P_s^3$, in the sense that $P_3^c(\cdot) = \pi(\cdot\times \calGsr)$ 
and $P_3^s(\cdot) = \pi(\calGcr \times \cdot )$. We refer to such joint distributions as \textit{couplings}. Consider the minimization problem
\begin{linenomath}\begin{equation}
\label{eq:kantorovich}
\calW(P_3^c, P_3^s) = \inf_{\pi \in \Pi(P_3^c, P_3^s)} \int_{\calGcr \times \calGsr} W(g, h) \ddi\pi(g,h).
\end{equation}\end{linenomath}
When the infimum in \eqref{eq:kantorovich} is achieved by a coupling $\pi_0$, 
this last is known as an \textit{optimal coupling}. 
When an optimal coupling is 
supported on a set of the form $\{(h, T(h)): h \in \calGcr\}$,
for some map $T:\calGcr \to \calGsr$, it can be seen that 
$T$ is in fact an optimal transport map between $P_3^c$ and $P_3^s$. 
The Kantorovich problem \eqref{eq:kantorovich} is therefore a relaxation 
of the Monge problem \eqref{eq:monge}. Unlike the latter, however, the 
minimization problem \eqref{eq:kantorovich} is always feasible since
$\Pi(P_3^c, P_3^s)$ is non-empty; indeed, $\Pi(P_3^c,P_3^s)$ always contains the independence coupling $P_3^c \otimes P_3^s$. 
Moreover, the infimum in the Kantorovich problem is achieved as long as the cost function $W$ is lower semi-continuous, 
and the measures $P_3^c$ and $P_3^s$ satisfy a mild moment condition (\cite{villani2008}, Theorem 4.1). 
We also note that the optimal objective value $\calW(P_3^c, P_3^s)$ defines a metric
between probability measures called the (first-order) Wasserstein distance~\citep{villani2003}, or Earth Mover's distance~\citep{rubner2000}.

Using the Kantorovich relaxation,  we   now formalize
Assumption~\ref{assm:ot_pre} into the following condition, which we shall
require throughout the remainder of this section.  
\begin{assumption}
\label{assm:ot}
Assume there exists an optimal   coupling $\pi_0 \in \Pi(P_3^c, P_3^s)$ between $P_3^c$ and $P_3^s$.
Given a pair of random variables $(H^c, H^s) \sim \pi_0$, let $\pi_0(\cdot|h)$ denote the 
conditional distribution of $H^s$ given $H^c=h$, for any $h \in \calGcr$. 
Then, the following implication holds:
\begin{linenomath}\begin{equation}
\label{eq:coupling_pushforward}
\begin{aligned}
 G^c &\sim P_4^c \\[-0.05in]
 G^s | G^c &\sim \pi_0(\cdot|G^c)
\end{aligned}~~ \Longrightarrow~~ G^s \sim P_4^s.
\end{equation}\end{linenomath}
\end{assumption}
Assumption~\ref{assm:ot} requires the $3b$ and $4b$ distributions to be sufficiently similar for 
their restrictions to the Signal and Control Regions to be related by a common conditional distribution. 
It further postulates that this conditional distribution is induced by the optimal coupling~$\pi_0$. 
Heuristically, $\pi_0(\cdot|H)$ plays the role of a multivalued optimal transport map  
for pushing an event $H$ from the distribution $P_3^c$ onto $P_3^s$. Assumption~\ref{assm:ot} requires this map to additionally push the distribution $P_4^c$ onto its counterpart $P_4^s$
in the Signal Region.  In the special case where
there exists an optimal transport map $T_0$ from $P_3^c$ to $P_3^s$, we note that
$\pi_0 = (Id, T_0)_\# P_3^c$ is an optimal coupling of $P_3^c$ with $P_3^s$, where $Id$ denotes the identity map. 
In this case, equation~\eqref{eq:coupling_pushforward} is tantamount to equation~\eqref{eq:pushforward_assm2'}.

\subsection{Background Estimation} \label{sec:BackgroundEstimation}
We next derive estimators for the background distribution $P_4^s$ under Assumption~\ref{assm:ot}.
It follows from equation~\eqref{eq:coupling_pushforward} and the law of total probability that
\begin{linenomath}
\begin{equation*}
P_4^s(\cdot) = \int_{\calG_c} \pi_0(\cdot|g) \ddi P_4^c(g).
\end{equation*}
\end{linenomath}
Since $\pi_0(\cdot|g)$ is  
the distribution of $H^s$ given $H^c = g$, induced by the optimal coupling $\pi_0$, it is an identified parameter
which can be estimated using only the $3b$ data.  
Given an estimator $\hat \pi(\cdot|g)$ of this quantity, 
and an estimator $\hat P_{4,\nfc}^c$ of $P_{4}^c$, 
it is natural to consider the plugin estimator of the background distribution $P_4^s$, given by
\begin{linenomath}\begin{equation}
\label{eq:general_ot_based_estimator}
\hat P_{4}^s(\cdot ) := \int_{\calG_c} \hpi(\cdot|g) \ddi\hat P_{4,\nfc}^c(g).
\end{equation}\end{linenomath}
In what follows, we begin by defining an estimator $\hpi(\cdot|g)$ in Section~\ref{sec:empirical_coupling}, 
followed by two candidates for the estimator $\hat P_{4,\nfc}^c$, 
leading to two distinct background estimation methods described in Sections~\ref{sec:ot_nn_estimator} 
and~\ref{sec:ot_fvt_estimator}.  
In Section~\ref{sec:ot_unnormalized}, we briefly discuss how these constructions also lead to estimators of the unnormalized
intensity measure $\binten_{4}^s$. We then provide discussion and comparison of these methodologies in Section~\ref{sec:ot_discussion}. 

\subsubsection{The Empirical Optimal Transport Coupling}
\label{sec:empirical_coupling}
A natural plugin estimator for the coupling $\pi_0$ is the optimal  coupling $\hat \pi$ between
the empirical measures $P_{3,\ntc}^c$ and $P_{3,\nts}^s$. In detail, denoting by $\hat q \in \bbR^{\ntc\times\nts}$
the joint probability mass function of $\hat\pi$, the empirical Kantorovich problem takes the following form:
\begin{linenomath}\begin{equation}
\label{eq:discrete_kantorovich}
\begin{multlined}[0.7\textwidth]
\hat q = (\hat q_{ij}) \in \argmin_{(q_{ij}) \in \bbR^{\ntc\times\nts}}
  \sum_{i=1}^{\ntc} \sum_{j=1}^{\nts}   q_{ij} W(H_i^c, H_j^s),
 ~~\text{s.t.} \  q_{ij} \geq 0, ~
					  \sum_{i=1}^{\ntc} q_{ij} = \frac 1 {\nts}, ~~
					  \sum_{j=1}^{\nts} q_{ij} = \frac 1 {\ntc}.
\end{multlined}					  
\end{equation}\end{linenomath}
Equation \eqref{eq:discrete_kantorovich} is a finite-dimensional linear program, 
for which exact solutions may be computed using network simplex algorithms such as the Hungarian algorithm \citep{kuhn1955a}.
We refer to~\cite{peyre2019} for a survey. 
We then define the estimator $\hat \pi (\cdot|H_i^c)$, for $i\in[\ntc]$, 
as the discrete distribution  
over $\{H_1^s, \dots, H_{\nts}^s\}$ with probability mass function  
\begin{linenomath}\begin{equation*}\hat q_{j|i} = \frac{\hat q_{ij}}{\sum_{k=1}^{\nts} \hat q_{ik}} =\ntc\cdot  \hat q_{ij},\quad j=1, \dots, \nts.\end{equation*}\end{linenomath}

We are now  in a position to define estimators of the background distribution $P_4^s$. 

\subsubsection{The \ot Estimator} 
\label{sec:ot_nn_estimator}
We first consider the general estimator in equation~\eqref{eq:general_ot_based_estimator} 
when $\hat P_{4,\nfc}^c$ is the empirical measure  
$P_{4,\nfc}^c$. This choice is perhaps most natural, but it requires us to
perform out-of-sample evaluations of the estimator $\hpi(\cdot|g)$.
Indeed, recall that the latter is defined over $\{H_1^c, \dots, H_{\ntc}^c\}$,
whereas $P_{4,\nfc}^c$ is supported on  $\{G_1^c, \dots, G_{\nfc}^c\}$. 

We extend the support  of $\hat\pi(\cdot|g)$ 
to all $g \in \calGcr$ 
  using a variant of the nearest
neighbors method for nonparametric regression~\citep{biau2015}. 
A similar procedure has also been used, for instance, by~\cite{flamary2021,manole2021c}.
Let $k \geq 1$ be an integer. 
For all $g \in \calGcr$, let $I_k(g)$ denote the indices
of the $k$-nearest neighbors of $g$ with respect to $W$, among
$H_1^c, \dots,H_{\ntc}^c$. Specifically, we set $I(g)=\{j_1, \dots, j_k\} \subseteq [\ntc]$
where
\begin{linenomath}
\begin{equation*}
W(g, H_{j_1}^c) \leq \dots\leq W(g, H_{j_k}^c) \leq W(g, H_j^c), \quad 
\text{for all } j \in [\ntc] \setminus \{j_1, \dots, j_k\}.
\end{equation*}
\end{linenomath}
Furthermore, define the inverse distance weights
\begin{linenomath}\begin{equation}
\label{eq:inverse_distance_weights}
\omega_i(g) = \frac{1/W(g, H_i^c)}{\sum_{l \in I_k(g)}1/W(g, H_l^c)}, \quad i\in I_k(g),
\end{equation}\end{linenomath}
with the convention $\infty/\infty=1$.
We then define for all $g \in \calGcr$,  
\begin{linenomath}\begin{equation}
\label{eq:hat_pi_NN}
\hat\pi_{k\mathrm{NN}}(\cdot|g) = \sum_{i\in I_k(g)} \omega_i(g) \hat \pi(\cdot|H_i^c).
\end{equation}\end{linenomath}
The estimator $\hat\pi_{k\mathrm{NN}}(\cdot|g)$ couples $g$ with all of the events to which its
$k$-nearest neighbors are coupled under $\hpi$. The coupling values
which correspond to the closest nearest neighbors are assigned higher weights $\omega_i(g)$.
Furthermore, we note that when $g \in \{H_1^c, \dots, H_{\ntc}^c\}$,
 it holds that 
$\hat\pi_{k\mathrm{NN}}(\cdot|g) = \hat\pi(\cdot|g)$. 
With these defintions, the generic estimator~\eqref{eq:general_ot_based_estimator} takes the   form, 
\begin{linenomath}
\begin{equation*}
\hat P_{4,k\mathrm{NN}}^s(\cdot) := \int_{\calGcr} \hat\pi_{k\mathrm{NN}}(\cdot|g) \ddi P_{4,\nfc}^c(g) = 
    \frac 1 {\nfc} \sum_{\ell=1}^{\nfc} \sum_{i\in I_k(G_\ell^c)} \omega_i(G_\ell^c) \hat \pi(\cdot|H_i^c),
\end{equation*}
\end{linenomath}
or equivalently,
\begin{linenomath}\begin{equation*}\hat P_{4,k\mathrm{NN}}^s = 
\frac {\ntc} {\nfc} \sum_{j=1}^{\nts}  \left(\sum_{\ell=1}^{\nfc} \sum_{i\in I_k(G_\ell^c)} \omega_i(G_\ell^c) \hat q_{ij} \right)\delta_{H_j^s}.\end{equation*}\end{linenomath}
We refer to $\hat P_{4,k\mathrm{NN}}^s$ as the OT-$k$NN (Optimal Transport--$k$ Nearest Neighbor) estimator of $P_4^s$.

\subsubsection{The OT-FvT Estimator}
\label{sec:ot_fvt_estimator}
The rate of production of $3b$ events typically exceeds that of $4b$ events 
by one order of magnitude (cf.\ Section~\ref{sec:simulation}). As a result, 
in the general formulation~\eqref{eq:general_ot_based_estimator} of our optimal transport map estimators, we expect to have access to a  smaller sample
size $\nfc$ for estimating the 
distribution~$P_4^c$, than the sample sizes $\ntc$ and $\nts$ for estimating the optimal transport coupling $\pi_0$. 
Motivated by this observation, we next define  an estimator $\hat P_{4,\nfc}^c$  which can leverage the larger $3b$ sample size $\ntc$.

Let $p_j^c = dP_j^c/d\nu_0$ denote the density of $P_j^c$ for $j=3,4$. Recall from 
Section~\ref{sec:classifier_method} that for any event $g$, $\psi(g)$ denotes the probability
that a random event $G$ arose from the $4b$ distribution as opposed to the  $3b$ distribution, given that $G = g$.
Furthermore, $\hat\psi(g)$ denotes the $[0,1]$-valued output of the FvT classifier
for discriminating $4b$ events from $3b$ events. 
Recall further
that for any $g \in \calGcr$, it holds that
$p_4^c(g)/p_3^c(g) = (\binten_3^c(\calGcr)/\binten_4^c(\calGcr))\cdot(\psi(g)/(1-\psi(g))$,
or equivalently, 
\begin{linenomath}
\begin{equation*}
P_4^c(A) = \frac{\binten_3^c(\calGcr)}{\binten_4^c(\calGcr)} \int_A \frac{\psi(h)}{1-\psi(h)} \ddi P_3^c(h), \quad A \in \bbB(\calGcr).
\end{equation*}
\end{linenomath}
We define a plugin estimator of the above quantity via
\begin{linenomath}\begin{equation}
\label{eq:P4c_ot_fvt_estimator}
\hat P_{4,\nfc}^c(A) = \frac{\ntc}{\nfc} \int_A \frac{\hpsi(h)}{1-\hpsi(h)} \ddi P_{3,\ntc}^c(h), \quad A \in \bbB(\calGcr).
\end{equation}\end{linenomath} 
$\hat P_{4,\nfc}^c$ can be viewed as a reweighted version of the empirical measure $P_{3,\ntc}^c$.
The weights are chosen to make the $3b$ sample resemble a $4b$ sample, by using the FvT classifier
to estimate the density ratio $p_4^c/p_3^c$. Since the $3b$ sample is one order of magnitude larger than the $4b$ sample, 
we heuristically expect this estimator to have smaller theoretical risk than the empirical measure $P_{4,\nfc}^c$
whenever the density ratio $p_4^c/p_3^c$ is smooth.

A second motivation for using the estimator $\hat P_{4,\nfc}^c$ is the fact that it is supported on
the domain of definition of the in-sample empirical optimal transport coupling $\hpi(\cdot|g)$. 
We therefore do not need to extend the domain of this estimator, unlike the previous section.
With these choices, the generic estimator in equation~\eqref{eq:general_ot_based_estimator} takes the following form: 
\begin{linenomath}\begin{equation} 
\label{eq:ot_fvt_estimator}
\hat P_{4,\mathrm{OF}}^s  := \int_{\calGcr} \hpi(\cdot|g) \ddi\hat P_{4,\nfc}^c(g)
=  \frac{\ntc}{\nfc} \sum_{j=1}^{\nts} \left( \sum_{i=1}^{\ntc} \frac{\hpsi(H_i^c)}{1-\hpsi(H_i^c)} \hat q_{ij}\right)\delta_{H_j^s} .
\end{equation}\end{linenomath}
We refer to $\hat P_{4,\mathrm{OF}}^s$ as the OT-FvT (Optimal Transport--Four vs. Three) estimator of $P_4^s$.  
 
\subsubsection{Estimation of the Background Normalization} 
\label{sec:ot_unnormalized} 
We briefly show how the OT-$k$NN and OT-FvT estimators 
can also be used to estimate the unnormalized background intensity function $\binten_4^s$. 
which requires the following assumption.
\begin{assumption}
\label{assm:abcd}
It holds that $\displaystyle \binten_4^s(\calGsr) = \binten_3^s(\calGsr)\binten_4^c(\calGcr)/\binten_3^c(\calGcr).$
\end{assumption}
Assumption~\ref{assm:abcd} implies that the ratio of the number of $4b$ to $3b$ events in the Control Region should be the same as 
that in the Signal Region. 
 Under this assumption, a natural estimator for $\binten_4^s(\calGsr)$ is simply given by
$\nfc\nts/\ntc$.  
Therefore, under Assumptions~\ref{assm:ot}--\ref{assm:abcd}, the probability measures $\hat P_{4,k\mathrm{NN}}^s$ and $\hat P_{4,\mathrm{OF}}^s$
can be used to define the following two estimators of the unnormalized background intensity measure $\binten_4^s$,
\begin{linenomath}\begin{align}
\label{eq:unnormalized_ot_estimators}
\hat\binten_{4,k\mathrm{NN}}^s = \frac{\nfc\nts}{\ntc} \hat P_{4,k\mathrm{NN}}^s,\quad
\hat\binten_{4,\mathrm{OF}}^s = \frac{\nfc\nts}{\ntc} \hat P_{4,\mathrm{OF}}^s.
\end{align}\end{linenomath}
We respectively refer to the above measures as the OT-$k$NN and OT-FvT estimators of $\binten_4^s$,
or simply as the OT-$k$NN and OT-FvT  methods.

\subsubsection{Remarks}
\label{sec:ot_discussion}

We summarize the three background estimation methods, FvT, OT-$k$NN and OT-FvT, in  Table~\ref{tab:compare_formulas}, 
and make the following remarks:
\begin{itemize}
\item Assumption~\ref{assm:ot} is the primary modeling assumption required
by OT-$k$NN and OT-FvT. We view this condition as being complementary to Assumption~\ref{assm:classifier}(ii), 
required by the FvT method. Indeed, it involves an extrapolation (of an optimal  coupling) from the $3b$ to $4b$ distribution, 
rather than an extrapolation (of a density ratio) from the Control Region to the Signal Region.

\begin{table}[t]

\caption{\label{tab:compare_formulas}Summary of the three background estimation methods: FvT, OT-$k$NN, and OT-FvT. 
The final estimator for each method takes the form
$\hat\binten_{4}^s \propto \sum_{j=1}^{\nts} v_j \delta_{H_j^s},$
for the values of $v_j$ listed in the table.
} 
\begin{tabular}{c|c|c|c}
$\begin{matrix}
\mathrm{Estimator}\\
\text{(of the form} \propto \sum_{j=1}^{\nts} v_j \delta_{H_j^s})
\end{matrix}$ & FvT  & OT-FvT& OT-$k$NN \\
\hline\hline
$v_j$  & $\displaystyle\frac{\hat \psi(H_j^s)}{1-\hat \psi(H_j^s)}$
	   & $\displaystyle\sum_{i=1}^{\ntc} \frac{\hpsi(H_i^c)}{1-\hpsi(H_i^c)} \hat q_{ij}$
	   & $\displaystyle\sum_{\ell=1}^{\nfc} \sum_{i\in I_k(G_\ell^c)} \omega_i(G_\ell^c) \hat q_{ij} $ 
\end{tabular}  
\end{table}

\item The OT-FvT estimator~\eqref{eq:ot_fvt_estimator} 
can alternatively be interpreted through the lens of 
domain adaptation
for the FvT classifier. To make this connection clear, 
suppose for simplicity that $\ntc = \nfc$.
In this case, it can be shown that $\hat\pi$ is in fact induced by an optimal transport map, in the sense
that there exists a permutation $\htau:[\ntc] \to [\ntc]$ such that 
\begin{linenomath}\begin{equation*}\hat q_{ij} = I(i = \htau(j))/\ntc, \quad i,j=1, \dots, \ntc.\end{equation*}\end{linenomath}
The FvT and OT-FvT estimators then take the following form:
\begin{linenomath}\begin{equation*}\hat \binten_{4,\mathrm{FvT}}^s \propto \sum_{j=1}^{\nts}  \frac{\hpsi(H_j^s)}{1-\hpsi(H_j^s)} \delta_{H_j^s},
\qquad \hat\binten_{4,\mathrm{OF}}^s \propto \sum_{j=1}^{\nts}  \frac{\hpsi(H_{\htau(j)}^c)}{1-\hpsi(H_{\htau(j)}^c)} \delta_{H_j^s} .\end{equation*}\end{linenomath}
While the FvT method evaluates the density ratio estimator $\hpsi/(1-\hpsi)$ at events $H_j^s$ in the Signal Region, 
the OT-FvT method evaluates it at the events $H_{\htau(j)}^c$ in the Control Region, to which the events $H_j^s$ are mapped
under the empirical optimal coupling $\hpi$. The OT-FvT method thus circumvents the evaluation of $\hpsi$ outside
the region where it was trained. 
Optimal transport has similarly been used in past literature 
as a tool for domain adaptation between train and test data in 
classification problems (cf.~Section~\ref{sec:lit}).

\item In defining the estimator OT-$k$NN, we proposed to extend the domain of definition 
of the empirical optimal transport coupling $\hat\pi(\cdot|g)$
to the entire space $\calGcr$ 
via nearest neighbor extrapolation; cf. equation~\eqref{eq:hat_pi_NN}. 
It was shown by~\cite{manole2021c} that, for the quadratic optimal transport
problem over Euclidean space, such a procedure has statistically minimax optimal risk for estimating 
the underlying optimal transport map $T_0$, assuming that it exists and is Lipschitz continuous. 
Nevertheless, the risk of this estimator suffers severely from the curse of dimensionality, and does not generally improve when $T_0$
enjoys higher regularity. \cite{manole2021c} and \cite{deb2021} have instead shown that plugin estimators
of $T_0$ based on density estimates of $P_3^c$ and $P_3^s$ may achieve improved convergence rates in such settings. 
In our context, it is challenging to perform 
density estimation over the space of measures $\calG$---and particularly
over the non-convex set $\calGcr$---thus we did not follow this approach. 
Our aim was instead to alleviate the curse of dimensionality  inherent to the OT-$k$NN method
by introducing the OT-FvT method. Indeed, we view the task of estimating $P_4^c$ as a larger statistical bottleneck 
than that of estimating $\pi_0$, and the estimator $\hat P_{4,\nfc}^c$ (used by the OT-FvT method) may potentially achieve smaller risk than the empirical measure~$P_{4,\nfc}^c$ (used by the OT-$k$NN method).

\item 
\cite{manole2021c} additionally show that the value $k=1$ suffices for the estimator $\hat\pi_{k\mathrm{NN}}$ to enjoy
optimal theoretical risk. 
In our work, we nevertheless allow for $k$ to be greater than 1 in order to leverage the larger size of the $3b$ sample. 
For example, when $k=1$, the estimator $\hat \binten_{4,k\mathrm{NN}}^s$ is supported on at most   $\nfc$ events, 
whereas it can be supported on as many as $\nts\gg \nfc$ events if $k$ is chosen sufficiently large. 
In practice, we recommend choosing $k$ to be as small as possible while ensuring that
$\hat \binten_{4,k\mathrm{NN}}^s$ has support size on the same order as $\nts$---this typically
amounts to choosing $k$ to be on the order of $\nts/\nfc$. 
In our simulation study (cf. Section~\ref{sec:simulation}), we therefore choose the value $k=10$, 
but also illustrate the performance of the OT-$k$NN method for other values of $k$.

\item We have chosen to separately estimate the probability measure $P_4^s$ and the normalization
$\binten_4^s(\calGsr)$, because
the classical optimal transport problem is only well-defined between measures with the same total mass. 
A possible alternative is to consider the {\it partial}~\citep{figalli2010} or
 {\it unbalanced}~\citep{liero2018}  optimal transport problems between
the unnormalized intensity measures $\binten_3^c$ and $\binten_3^s$. These variants of  optimal transport
are well-defined between measures that have possibly different mass, but have the downside of introducing tuning parameters.
As we explain in Section~\ref{sec:simulation},  the normalizations $\binten_3^c(\calGcr)$ and $\binten_3^s(\calGsr)$ 
 are of the same  order of magnitude, and can in fact be made
to coincide by tuning the definition of the Control and Signal regions, 
thus we have simply focused our attention on the classical (balanced) optimal transport problem in this work. 
Nevertheless, in the following subsection, we will employ
a variant of the partial optimal transport problem
to define the metric $W$.

\end{itemize}

\subsection{A Metric between Collider Events}
\label{sec:ot_metric}
\begin{figure}[t]
\includegraphics[width=.46\textwidth]{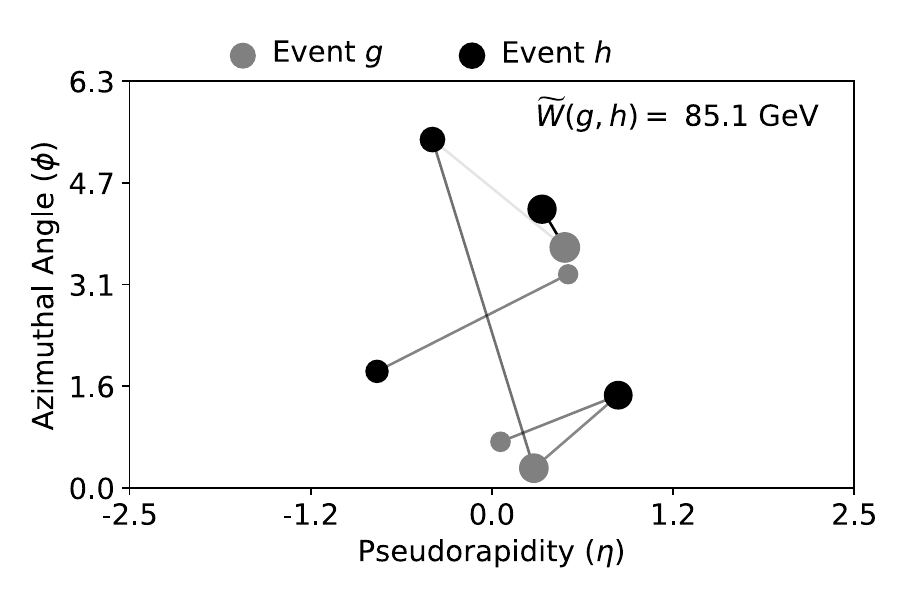}
\includegraphics[width=.46\textwidth]{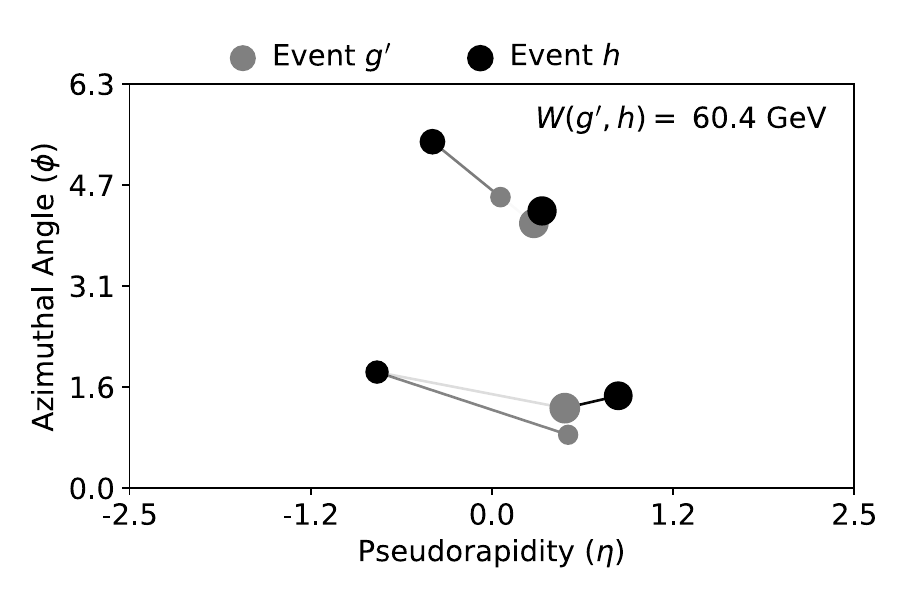}
\caption{\label{fig:emd} Left: $(\eta,\phi)$-plot of two events $g, h \in \calG$. Each point represents
a constituent jet, with size proportional to its $p_T$ value. A line connecting
the $i$-th jet of event $g$ to the $j$-th jet of event $h$ indicates a nonzero
value of the optimal coupling $f_{ij}$, with line darkness increasing as a function of the magnitude of $f_{ij}$. 
Right: $(\eta,\phi)$-plot of events $g', h \in \calG$, where $g' \simeq g$ is an approximate minimizer in Eq.\,\eqref{eq:final_W}. 
One has   $W(g, h) = W(g', h)<\widetilde W(g, h)$. }
\end{figure}

We now describe a candidate for 
the metric $W$ on $\calG$.
Recall that the Kantorovich problem in \eqref{eq:kantorovich} gave rise to the
Wasserstein distance $\calW$ between probability distributions over $\calG$. 
By a recursion of ideas, we will also define  $W$ to be a Wasserstein-type metric,  arising from the optimal transport problem
between constituent jets of events.
This approach was introduced by \cite{komiske2019}. They propose to metrize $\calG$ using a variant of the Wasserstein distance which is well-defined between
measures with non-equal mass~\citep{peleg1989,pele2008}. 
Given any two collider events
$g = \sum_{j=1}^4 p_{T_j} \delta_{(\eta_j, \phi_j, m_j)} \in \calG$,
$h = \sum_{j=1}^4 p'_{T_j} \delta_{(\eta_j', \phi_j', m_j')} \in \calG$, 
the metric is defined by
\begin{linenomath}\begin{align}
 \nonumber
\widetilde W(g, h)
 = \min_{(f_{ij}) \in \bbR^{4\times 4}} \frac 1 R \sum_{i=1}^4& \sum_{j=1}^4 f_{ij} \sqrt{(\eta_i - \eta_j')^2 + (\phi_i - \phi_j')^2} + \left|\sum_{i=1}^4 (p_{T_i} - p'_{T_i})\right| \\
\label{eq:emd}
   \text{s.t.} \quad  f_{ij} \geq 0, \quad \sum_j& f_{ij} \leq p_{T_i}, \quad
    								 \sum_i f_{ij} \leq p_{T_j}', \quad \sum_{i,j} f_{ij} = \min(\sum_i p_{T_i}, \sum_j p_{T_j}'),
\end{align}\end{linenomath}
for a tuning parameter $R > 0$. We make several remarks about this definition.
\begin{itemize}
\item In the context of particle physics, the coupling  $f_{ij}$
is naturally interpreted as a flow of energy (measured
in terms of the transverse momentum $p_T$) from jet $i$ of $g$ to jet $j$
of $h$, as depicted in Figure \ref{fig:emd}. $\widetilde W(g,h)$ thus measures the smallest possible transport of energy required to rearrange the jets of the
event $g$ into those of $h$.
\item We have followed~\cite{komiske2019} by omitting the mass variables $m_j$ and $m_j'$
from the definition of $\widetilde W$. This choice is further discussed in 
the context of our simulation study in Section~\ref{sec:simulation}.
\item The tuning parameter $R$ trades-off 
the influence of the angular variables $\phi_i,\eta_i$, and that
of the energy variables $p_{T_i}$. Our choice 
of $R$ is further discussed in~Section~\ref{sec:simulation}. 
\end{itemize}

The metric $\widetilde W$ does not, however, take into account the 
equivalence relation $\simeq$ over $\calG$ defined in equation~\eqref{eq:equivalence}. 
For example, $\widetilde W(g,h)$ could be nonzero even when $g$ and $h$ are deemed equivalent 
for our purposes. 
We therefore define our final metric $W$ by  
\begin{linenomath}\begin{align}
\label{eq:final_W}
W(g,h)
 = \inf\Big\{\widetilde{W}(g', h):   g' \simeq g, \ g' \in \calG \Big\}, \quad g, h \in \calG.
\end{align}\end{linenomath}
Strictly speaking, $W$ now becomes a metric over the set of equivalence classes
of events induced by $\simeq$. We refer to Figure~\ref{fig:emd} for an illustration. 
In practice, we numerically approximate $W$ using a procedure described in Appendix~\ref*{app:emd}.

\section{Simulation Study}
\label{sec:simulation} 

\subsection{Simulation Description}
In this section, we compare the performance of the three background modeling methods \dfvt, \ot and FvT, 
on realistic simulated collider data, generated using
the MadGraph particle physics software \citep{alwall2011}. Code for reproducing 
this simulation study is publicly available\footnote{\texttt{\href{https://github.com/tmanole/HH4bsim}{https://github.com/tmanole/HH4bsim}}}.

Since $b$-tagging is imperfect, 
in practice, we expect the $3b$ and $4b$ samples to be composed
of a mixture of different multijet scattering processes which do necessarily arise from $b$-quarks. 
 We perform a
study in MadGraph to estimate
the relative scale of  
such processes. 
Assuming a $b$-jet tagging efficiency of 75\%, a charm jet tagging efficiency of
15\% and a light jet tagging efficiency of 1\%, we find that the $4b$ (resp. $3b$) sample consists 
of 90\% (10\%) events in a final state with four $b$ quarks, 7\% (9\%) events in a final state with two $b$ quarks and two charm quarks, 
and 4\% (80\%) in a final state with two $b$ quarks and two light quarks. In particular, we stress that a fraction 
of the $3b$ sample consists of mislabelled $4b$ events, which could be signal events. This signal contamination
is expected to be sufficiently small to be considered negligible for purposes of a signal discovery analysis, as in this paper.

We generate four-quark events in MadGraph according to the percentages listed above.~The calorimeters 
in the CMS detector are not perfect, and the measured jet energies have a finite resolution.
The distribution of the observed smeared energy is well-approximated by 
the normal distribution $N(E, \sigma^2(E))$, where $E$ denotes the true energy of a jet, and 
$\sigma(E)$~satisfies
\begin{linenomath}\begin{equation*}\left(\frac{\sigma(E)}{E}\right)^2 = \left(\frac{S}{\sqrt E}\right)^2  + \left(\frac N E \right)^2 + C^2,\end{equation*}\end{linenomath}
for some constants $S, N, C \geq  0$.
We apply this smearing to the quark four-vectors, setting $S = 0.98, N = 0, C = 0.054$.
For simplicity we set the quark masses to zero and omit~them from the 
the metric $W$.
When applying these methods to real data, it may, however, be useful to incorporate
the jet masses into the definition of $W$.
We also apply jet-level scale factors to account for the $p_T$ dependence of CMS $b$-tagging for light, charm and bottom quark~jets:
\begin{linenomath}\begin{equation*}
  \text{Scale Factor} = \begin{cases}
    (2.5\,  p_T\, e^{ -7\, p_T} + 0.6 )/0.75 &\text{$b$-quark} \\
    (       p_T\, e^{-10\, p_T} + 0.2 )/0.15 &\text{$c$-quark} \\
    (0.03\, p_T                 + 0.01)/0.01 &\text{$u,d,s$-quark or gluon},
  \end{cases}
\end{equation*}\end{linenomath}
where $p_T$ is measured in TeV. Events are weighted by the product of the scale factors for the $b$-tagged jets.

Following this pre-processing of the data, the pairing algorithm described in 
Section~\ref{sec:setup_background} is applied to all events, and those falling within the Control and Signal Regions
are kept. We define these regions according to equations~(\ref{eq:signal_region_defn}--\ref{eq:control_region_defn}), with the parameters $\sigma_c = 1.03$, $\kappa_s = 1.6$
and {\color{revs}$r_c = 30$\,GeV}.
The final sample consists of $\nts = 201,568$ events in the $3b$ Signal Region, 
$\ntc=159,427$ events in the $3b$ Control Region, 
$\nfs = 28,980$ events in the 4b Signal Region, 
and $\nfc = 22,053$ events in the 4b Control Region.
The order of magnitude of these sample sizes, as well as the proportion of $3b$ to $4b$ events, is similar to 
those used in recent di-Higgs analyses at the LHC \citep{atlas2019}. We also 
simulate a separate 4b sample of size approximately $10(\nts+\nfs)$, which we choose not to contain any signal events,
and whose distribution we
treat as the ground truth for the purpose of validating our background models. 

We additionally generate a Monte Carlo sample from the  SM di-Higgs signal distribution, with which 
the signal intensity rates $(S_j)_{j=1}^J$, used
to form  
the likelihood function~\eqref{eq:likelihood}, can be specified. For the purpose
of validating our background models, we train a $[0,1]$-valued classifier $\hat \xi$ (abbr. Signal vs. Background, or SvB, classifier)
to discriminate the $4b$ data from the Monte Carlo di-Higgs sample. Given that our simulated 4b sample contains no signal events, 
$\hat \xi$ forms a reasonable proxy for the theoretical binning function $\xi$ in equation~\eqref{eq:signal_classifier_setup}.  
The SvB classifier has the same architecture as that of the FvT classifier described in Appendix~\ref*{app:classifier_description}.
In the sequel, we refer to $\hat \xi(g)$ as the SvB value corresponding to an event $g$. 

Finally, we   discuss our choice of the parameter $R$ arising in the definition of the metric~$W$. 
In order for the two terms in the definition of $\widetilde W$
to be of comparable order, we make the requirement that $R$ lie within the range of the first summand in equation~\eqref{eq:emd}. 
We identify this range as follows. 
Since  $b$-tagging is only performed for values of $\eta$ lying in the interval $[-2.5,2.5]$, we impose $R \leq \sqrt{\pi^2+5^2} \approx 5.9$.
Furthermore, jet clustering algorithms used by CMS merge particles whose $(\eta,\phi)$-Euclidean distance is within $0.4$~\citep{cacciari2008,cms2017}, 
thus we impose $R \geq 0.4$. Now, since we expect that the largest
discrepancies between the Control and Signal Region distributions arise in the kinematic variables $(\eta,\phi)$, 
we choose the smallest possible value $R = 0.4$ when fitting the empirical optimal transport coupling $\hat\pi$. 
On the other hand, for the nearest-neighbor lookup of the OT-$k$NN method, we set $R=2.75$, which is the midpoint
of the interval $[0.4,5.9]$. We make no attempt to tune these values of $R$, and we leave open the question of
choosing them in a data-driven fashion. We compute the metric $W$ in part using the EnergyFlow Python library~\citep{energyflow}, 
and we compute optimal couplings between distributions of collider events using the Python Optimal Transport library~\citep{flamary2021}---see
Appendix~\ref*{app:ot_computation} for further details.

\subsection{Simulation Results}
\begin{figure}[t] 
\includegraphics[width=0.495\textwidth]{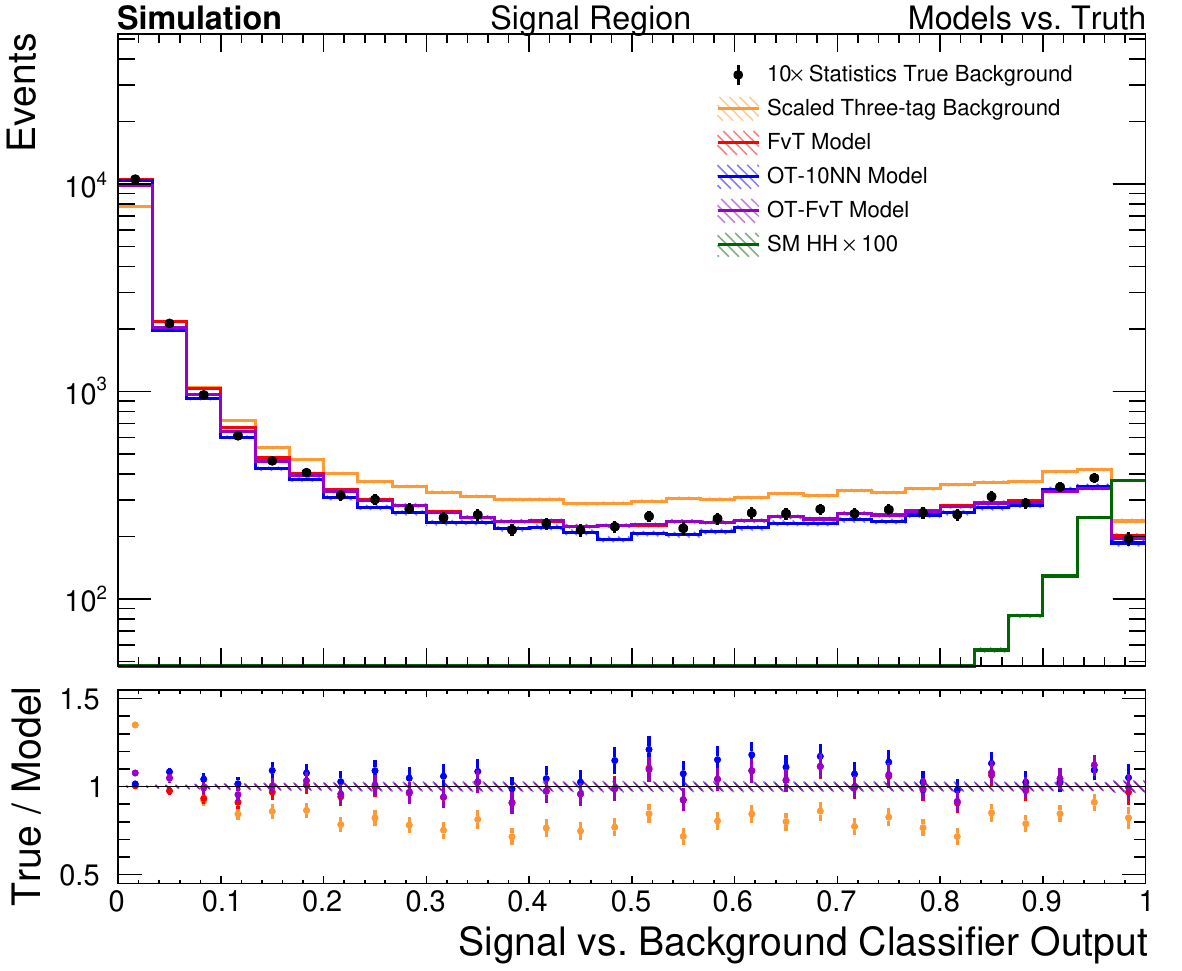}
\includegraphics[width=0.495\textwidth]{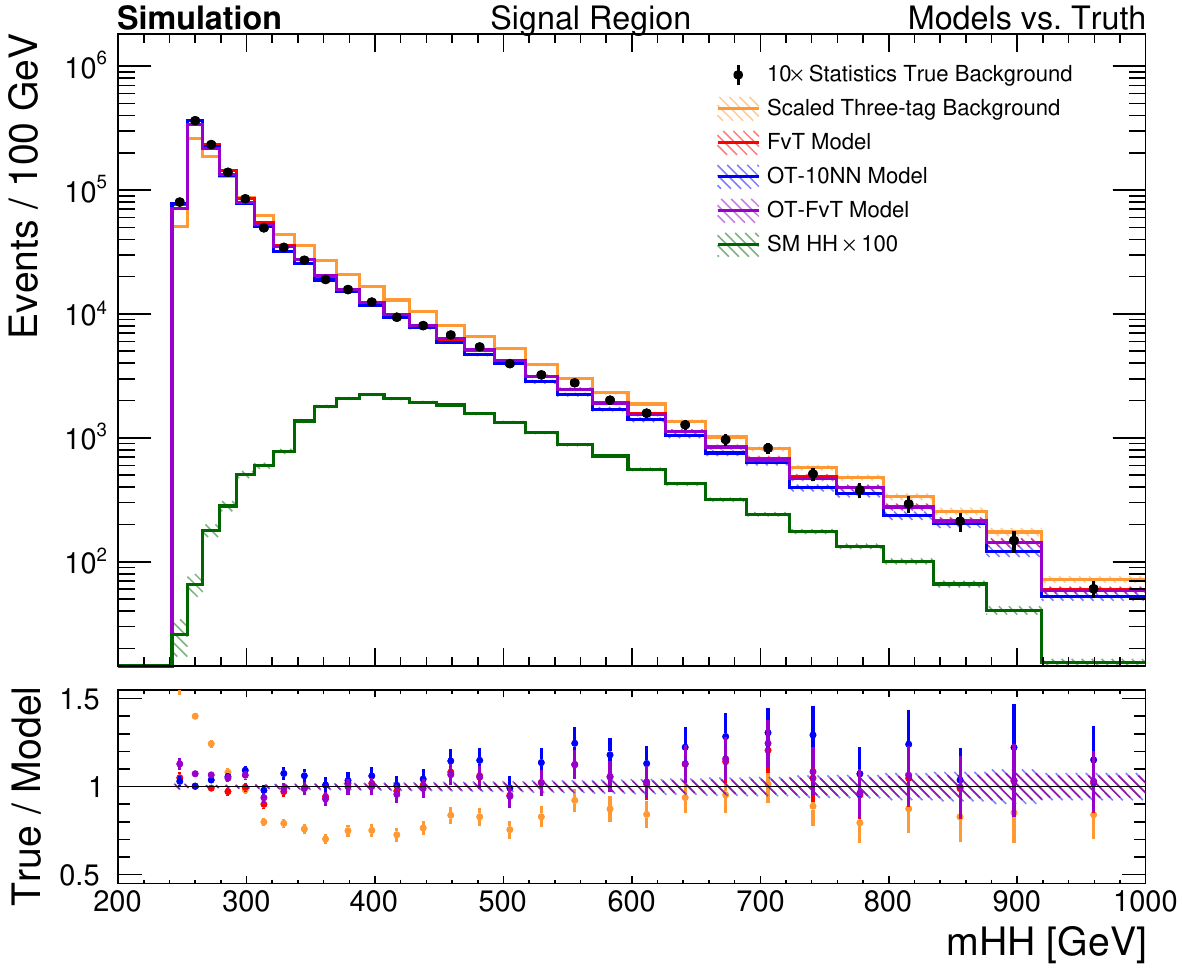} 
\caption{\label{fig:sim_SvB_log} Histograms of the the SvB classifier output variable (left) and the $m_{\mathrm{HH}}$ variable (right) for
the three background models as well as the upsampled 4b data (treated as the ground truth), 
the 3b data (normalized by the factor $\nts\nfc/\ntc$), and 
the di-Higgs signal sample (SM HH). Error bars in the $k$-th bin of any histogram denote $\pm \sqrt {N_k}$, where $N_k$ is the number of 
events per bin. Error bars in the ratio plot denote $\pm \sqrt {N_{k}}/N_{0k}$, where $N_{0k}$ is the number
of observed $4b$ events per bin. The dashed lines in the ratio plot denote $\pm \sqrt {1/N_{0k}}$.}
\end{figure}

The fitted 
 intensity measures $\hat\binten_{4}^s$ produced by the three background methods (FvT, OT-$k$NN and OT-FvT)
 are binned and plotted in Figure~\ref{fig:sim_SvB_log}. 
Logarithmic scales are used to better visualize 
signal-rich regions. Plots with additional kinematic variables are given in Appendix~\ref*{app:additional_simulation_results}. 
It can be seen that the three methods yield qualitatively similar estimates 
of the SvB intensity function. We recall that the SvB variable is of primary interest to model, as it is used
as the final discriminant when testing the signal hypothesis~\eqref{eq:hypotheses}. 
The $m_{\mathrm{HH}}$ variable
has also been used as the final discriminant in recent di-Higgs studies~\citep{bryant2018}.
Given an event $g \in \calG$
with dijet pairing $\{g^1,g^2\}$,
its $m_{\mathrm{HH}}$ value is 
 defined as follows\footnote{Equation~\eqref{eqn:mHH} can be interpreted as the four-body invariant mass after the dijet four-vectors have been corrected to have the Higgs boson mass.}, using the notation of Section~\ref{sec:setup_background},
\begin{linenomath}\begin{equation}
  \label{eqn:mHH}
  m_{\text{HH}}(g) = m\bigg( \frac{m_{\text{H}}}{m(g^1)}g^1 + \frac{m_{\text{H}}}{m(g^2)}g^2 \bigg).
\end{equation}\end{linenomath}
Once again, we observe that this variable is well-modelled by all three methods. 
Among the various kinematic variables which we analyzed, the ``$\Delta R_{jj}$--Other'' variable, appearing 
in Figure~\ref{fig:sim_mHH_dRjjOther_log} of Appendix~\ref*{app:additional_simulation_results}, presents one of the largest qualitative discrepancies between the three methods, and appears to be best-modelled by the FvT method. 
In all cases,  the three methods provide a significant
improvement compared to the uncorrected $3b$ sample.

In order to provide a quantitative comparison of these methods, 
we develop a heuristic two-sample test for testing equality between the
distribution of the fitted background models and of the true upsampled $4b$ data. 
Specifically, we form a proxy for a two-sample test by training classifiers to discriminate each of the background estimates
from the upsampled $4b$ data {\color{revs}(similar
approaches have previously been used in the high energy 
physics literature by \cite{krause2023,krause2023a}).}
For each classifier, we record the area under the receiver
operating characteristic curve (AUC; \cite{hanley1982}), and any deviation of this quantity from .5 is an indication
of mismodeling. We again choose our classifiers to 
be residual neural networks with the
architecture described in Appendix~\ref*{app:classifier_description}. Although this choice
is inherently favourable to the FvT method, and to some extent the OT-FvT method, we use it 
because it   coincides with the SvB classifier architecture, and will thus be most powerful at
detecting mismodeling in the features which are relevant for the final signal analysis.
{\color{revs} Another caveat with the use of the AUC as a performance metric is 
the fact that it quantifies the overall quality of the background estimates over 
the whole Signal Region, and might hence not be sensitive to local deviations in 
the most signal-rich part of the Signal Region. For this reason, it is necessary to 
also look at the quality of the estimates in terms of metrics that are localized to 
the most signal-rich part of the phase space, as is done in Figure~\ref{fig:sim_SvB_log} (left) for the 
di-Higgs signal. If the background estimates are to be used with any other alternative 
signal model, similar checks would need to be done with metrics localized for those signals. }
\begin{figure}[t]
\includegraphics[width=0.4\textwidth]{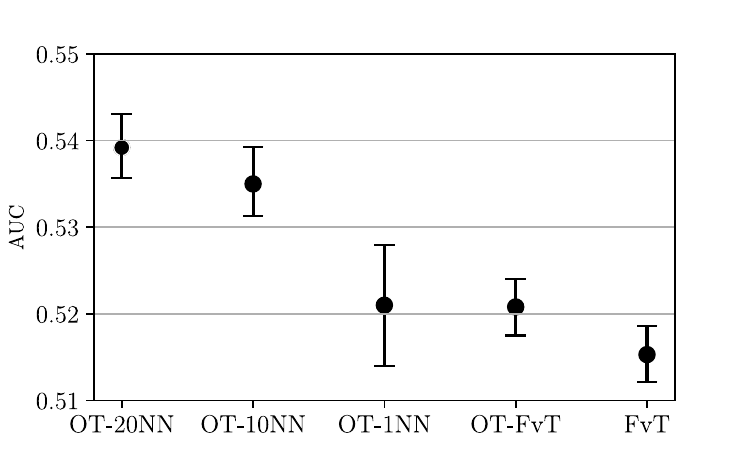} 
\caption{ 
\label{fig:aucs}
  Fitted AUC values obtained by discriminating each background model from the upsampled 4b data
using the FvT classifier, {\color{revs}in the Signal Region}, 
together with 95\% percentile bootstrap variability intervals, obtained by bootstrapping the predicted 
classifier probabilities. 1,000 bootstrap replications are used. Note that this bootstrap
procedure does not take into account the variability of the background estimators themselves.
 For the $3b$-tagged data, we obtain the AUC 0.5843, with variability interval [0.5812,0.5874].}
\end{figure}
 
\begin{figure}[H] 
\includegraphics[width=0.49\textwidth]{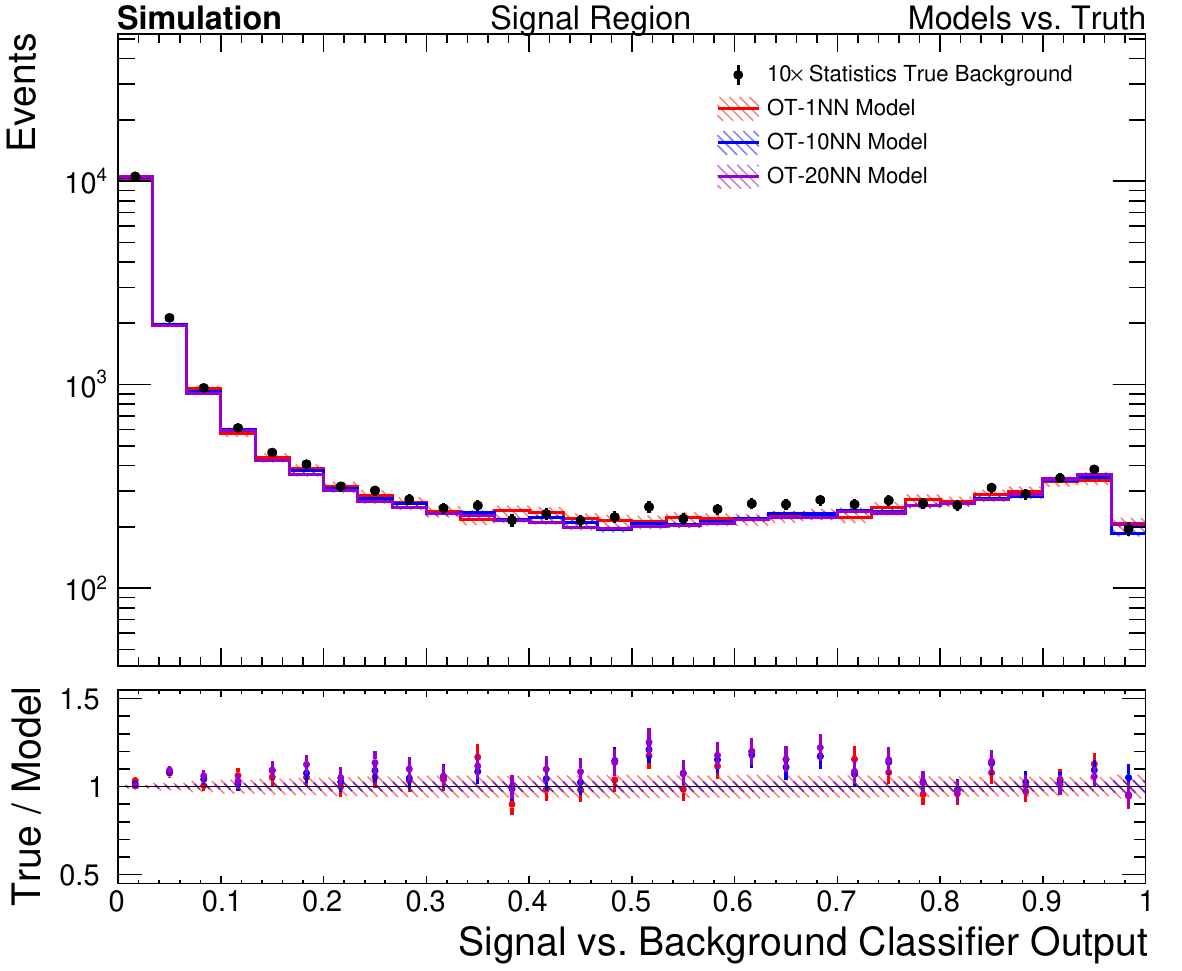} 
\includegraphics[width=0.49\textwidth]{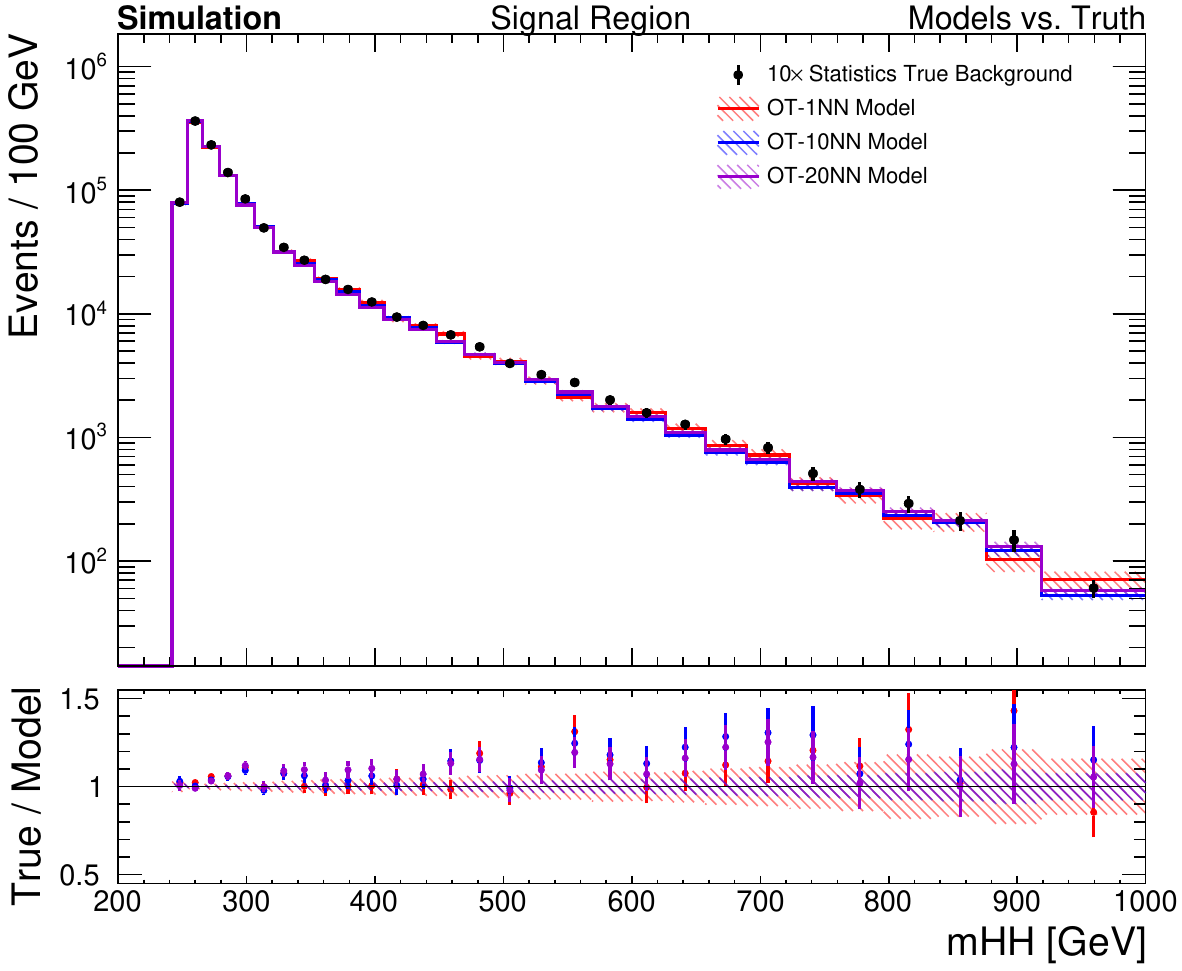} 
\caption{\label{fig:nn_comparison} Histograms of the SvB classifier output variable (left)
and the $m_{\mathrm{HH}}$ variable (right) for
the OT-$k$NN estimator, with $k\in \{1, 10, 20\}$. } 
\end{figure}

The fitted AUC value for each method is reported in Figure~\ref{fig:aucs}. 
Though all AUCs are significantly greater than .5, 
they are substantially lower for our background models
than for the benchmark method consisting of the uncorrected 3b sample. 
The FvT method has the lowest AUC, followed closely by the OT-FvT method and
OT-$k$NN method. While the OT-1NN method has comparable AUC point estimate as the OT-FvT method, we
emphasize that its variability interval is wider, which could have been anticipated from
the discussion in Section~\ref{sec:ot_discussion},
where we emphasized that the support size of $\binten_{4,1\mathrm{NN}}^s$
can be an order of magnitude smaller than $n_s$. In contrast, 
the OT-10NN and OT-20NN estimators have narrower variability intervals
than OT-1NN, but have markedly
larger AUC point estimates than the remaining methods. The performance of the OT-$k$NN method
for varying values of $k$ is also illustrated in Figure~\ref{fig:nn_comparison}
 as a function of the SvB and $m_{\mathrm{HH}}$ variables.

\begin{figure}[t] 
\includegraphics[width=0.9\textwidth]{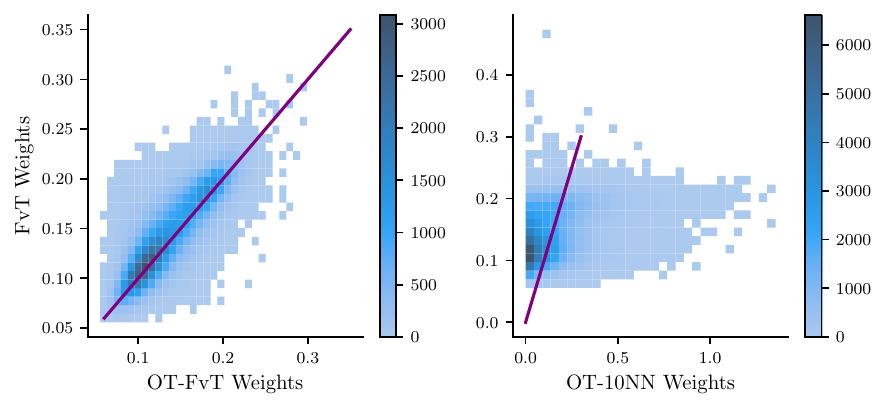}
\vspace{-0.25in}
\caption{\label{fig:weight_comparison}Bivariate histogram of the $3b$ data $H_1^s, \dots, H^s_{\nts}$ 
in the Signal Region, plotted in terms of the weights of the OT-FvT method against those of the FvT method (left), 
and of the OT-10NN method against those of the FvT method (right). The purple lines denote  the identity function.}
\end{figure}

We next provide a qualitative comparison of the fitted weights produced by the three background modeling methods.
Recall that these methods all take the form
\begin{linenomath}\begin{equation*}\hat\binten_4^s = \sum_{i=1}^{\nts} v_i \delta_{H_i^s},\end{equation*}\end{linenomath}
for some nonnegative weights $v_i$, which are summarized up to normalization in Table~\ref{tab:compare_formulas}. 
In Figure~\ref{fig:weight_comparison}, we plot the weights of the two optimal transport methods against 
those of the FvT method. We 
observe that the FvT and OT-FvT methods produce weights which are concentrated and symmetric around the identity. 
This implies that the odds ratio of the FvT classifier at a point $H_j^s$ in the Signal Region
behaves similarly to the odds ratio at any point $H_{i}^c$ 
in the Control Region to which $H_j^s$ is optimally coupled. 
This suggests that the transfer learning of the FvT classifier from the Control Region to the Signal Region is,
to some extent, well-modelled by the optimal transport coupling $\hat\pi$.
This observation heuristically suggests
that Assumptions~\ref{assm:classifier}--\ref{assm:ot} both hold in this simulation. 
In contrast to the method OT-FvT, we observe that the method OT-10NN produces markedly different weights than the FvT method, 
which can partly be anticipated from the discrete nature of the nearest neighbor extrapolation. 
We conjecture 
that the nearest-neighbor estimator of the optimal transport coupling has poorer theoretical risk than its counterpart
in the OT-FvT method.

 \section{Conclusion and Discussion}
\label{sec:conclusion}
Our aim has been to study the problem of data-driven background estimation, motivated by the ongoing search
for double Higgs boson production in the $4b$ final state. After recalling a widely-used approach to this problem based on
transfer learning of a multivariate classifier, our first contribution was to develop
the FvT classifier architecture which is tailored
to collider data, and which can serve as a powerful tool for implementing this methodology. 
Our primary contribution was then to propose a distinct background estimation method based on 
the optimal transport problem. 
A recurring theme throughout our work has been the complementarity of the modeling assumptions
made by these two distinct approaches, which allows them to be used as cross-checks for one another in practice.
We substantiate this point with a realistic simulation study, in which these methodologies appear to give 
consistent results despite their inherently distinct derivations.

Quantifying the uncertainty of our background estimates is a challenging problem left open by our work,
which is nonetheless crucial for applying our methods in practice.
In the experimental particle physics community, it is commonplace to measure both {\it statistical} uncertainties---those arising
from fluctuations of the data generating process---and {\it systematic} uncertainties---those 
arising from potential mismodeling~\citep{heinrich2007}. 
Both of these forms of uncertainty are challenging to quantify in our context. 
For instance, a prerequisite for quantifying the statistical uncertainty of the methods OT-$k$NN or OT-FvT
is to perform statistical inference for optimal transport maps. This is 
a difficult open problem in the statistical optimal transport literature, 
that has only been addressed for some special cases~\citep{rippl2016,ramdas2017},
and for regularized variants of optimal transport maps which differ from those used in our work~\citep{klatt2020,gunsilius2021, sanz2022,goldfeld2022}. 
The question of quantifying systematic uncertainties is more open-ended, and typically involves heuristics
for assessing the extent of potential mismodeling by the background estimation methods. Due to the complementarity of  
assumptions placed by our methods, any lack of closure between them could potentially play a role in quantifying their
systematic uncertainties. While further investigation is required to make such a proposal formal, it is our hope that
the optimal transport methodology presented in our work can help contribute to the challenging question of systematic
uncertainty quantification in the search for di-Higgs boson production, or in other searches at the Large Hadron Collider.

\begin{acks}[Acknowledgments]
	
We are grateful to the CMU Statistical Methods for the Physical Sciences (STAMPS) research group for insightful discussions and feedback throughout this work. 

\end{acks}

\begin{funding} This work was supported in part by NSF PHY-2020295 and NSF DMS-2053804.
TM was supported in part by the Natural Sciences and Engineering Research Council of Canada (NSERC), 
through a PGS D Scholarship.
\end{funding}

\appendix

\section{Translation for the High Energy Physicist}

\label{app:hepexTranslation}
The main body of the text has been primarily written with the statistics community in mind.
This appendix aims to bridge the gap between the language and formalism used by statisticians and that common in high-energy physics (HEP).

\textbf{Section~\ref{sec:signalSearches}} describes the standard HEP search formalism in the statistician's nomenclature.
We have a dataset $F$ consisting of background and some \textit{a priori} unknown amount of signal, parameterized by signal strength $\mu$.
The search is done in bins of a discriminating variable $\xi$, 
 which will---later in the text---be taken to be the output of a multivariate classifier trained to separate signal and background.
The challenge is to predict the amount of background in each of the classifier output bins.

The discussion relating to Borel measures is a formality that can be safely ignored by less technical readers.
These technicalities are relevant because our discussion of the optimal transport problem, below, is more naturally formulated in terms of distributions over sets of events rather than probability density functions.
Distributions are restricted to the Borel $\sigma$-algebra to avoid theoretically possible pathologies, which are unlikely to be of practical concern to the data analyst.

``Intensity measures'' have a normalization given by the expected number of events, whereas ``probability distributions'' are normalized to unity.
For example,  $\binten_4$ is the intensity of background events, with $\binten_4(\calG)$ the expected total number of these events. 
Later on, we will also use the notation $b_4$ to denote an ``intensity function'', which can be viewed as the unnormalized
probability density function corresponding to the intensity measure $\binten_4$.

\textbf{Section~\ref{sec:setup_background}} introduces the three-tag dataset $T$.
The three-tag data, when normalized to the number of four-tag events, provides a zeroth order estimate of the $4b$ background.
The main contribution of our work is in deriving corrections to the $3b$ data to better approximate the true $4b$ background.

An algorithm is used to pair jets into Higgs candidates and Signal and Control regions are defined using the reconstructed masses.
The ``empirical estimators of the measures $\binten_3^c, \binten_3^s, \binten_4^c$'' are effectively just the observed datasets in the respective regions.

\textbf{Section~\ref{sec:classifier_method}} describes a data-driven background estimation method used frequently   in HEP~\citep{Aaboud_2018,CMS4b2022}.
This method can be thought of as an extension of the ``ABCD method''.

Equation~\eqref{eq:ratio_classifier_density} points out that a probabilistic classifier ($\psi$) trained to separate $4b$ and $3b$ events can be used to define a function that returns the relative probability density ratios between $4b$ and $3b$ events. 
This density ratio can then be used to weight the $3b$ data to give an improved estimate of the $4b$ background.

We train a classifier, referred to as the ``FvT classifier'', to separate $4b$ and $3b$ data in the Control Region (CR).
Assumption~\ref{assm:classifier} formalizes the usual assumption that the classifier trained with events in the CR can be safely extrapolated to also weight $3b$ events in the Signal Region (SR).
With this assumption the predicted background in the SR can be obtained by weighting $3b$ SR events by the density ratio estimated in the CR, equation~\eqref{eq:FvTestimator}.

\textbf{Section~\ref{sec:ot_method}} presents a novel method for implementing the ABCD method across phase space.
Instead of extrapolating between the $3b$ and $4b$ samples---assuming the extrapolation is the same in the CR and SR---we propose extrapolating between the CR and SR---assuming the extrapolation is the same in the $3b$ and $4b$ samples.
The extrapolation across phase space is complicated because the CR and SR distributions are kinematically disjoint.
We cannot use a classifier to correct kinematic differences between the samples; any classifier trained on kinematically disjoint samples would achieve perfect separation, and the corresponding weights would be undefined.
Instead, we assume that the ``optimal transport'' map which maps events in the CR to the SR is the same for the $3b$ and $4b$ events.
The optimal transport (OT) map is the transformation that minimizes the ``cost'' in deforming a distribution supported in the CR into one
supported in the SR.
The idea is to find the optimal map between the $3b$ CR and $3b$ SR events and then ``apply the map'' to the $4b$ CR events.

Neither finding, nor applying the OT map is straightforward; Sections~\ref{sec:OTProblem} and~\ref{sec:empirical_coupling} explain how the OT map is estimated and Sections~\ref{sec:ot_nn_estimator} and~\ref{sec:ot_fvt_estimator}  presents two alternative methods for applying the map to $4b$ CR events.
In order to define the OT problem, we need a notion of cost, or distance, between events; our choice is presented in Section~\ref{sec:ot_metric}.

\textbf{Section~\ref{sec:OTProblem}} describes complications associated with estimating the OT maps with collider data.
Formally, the OT problem is known to have a unique solution on continuous distributions with a Euclidean cost function; however, we have a non-Euclidean cost function (see Section~\ref{sec:ot_metric}) and only have access to finite datasets drawn from the relevant probability distributions.  
So instead of solving for the OT map, we estimate a ``coupling'' between datasets that converges to the OT map in the infinite statistics limit.
This coupling is the solution of the ``Kantorovich optimal transport problem'' defined in equation~\eqref{eq:kantorovich} and is a (weighted) mapping of events in the $3b$ CR to events in the $3b$ SR.
Assumption~\ref{assm:ot} 
formalizes the assumption that the map fit with the $3b$ data is applicable to the $4b$ events. 

\textbf{Section~\ref{sec:empirical_coupling}} explains that the coupling is estimated
using the observed $3b$ CR and SR events.
Note that this coupling map is only defined for $3b$ CR events, whereas we need to apply it to $4b$ CR events to get a background estimate.
We have two different solutions for applying the OT map ``out-of-sample'' to $4b$ events, which are described in Sections~\ref{sec:ot_nn_estimator} and~\ref{sec:ot_fvt_estimator}. Both of these methods rely on the fact that $3b$ and $4b$ CRs share the same support.

\textbf{Section~\ref{sec:ot_nn_estimator}} presents the first solution (called OT-$k$NN), in which $4b$ CR events are mapped to the ``nearest'' $3b$ CR events.
The OT coupling of these nearest neighbors is then used to map the $4b$ CR event to the SR.
We can use an arbitrary number of nearest neighbors~($k$), where the neighbors are weighted based on how close they are to the input $4b$ CR event.
Again, in order for this to make sense we need a notion of distance between events, and we use the definition presented in Section~\ref{sec:ot_metric}.

\textbf{Section~\ref{sec:ot_fvt_estimator}} presents the second solution (called OT-FvT) for applying the OT map to $4b$ events.
Here, the $3b$ CR events are first weighted to look like the $4b$ CR events using the FvT classifier.
Note that, in this case, the FvT-weights are applied in-sample and thus no assumption of extrapolation is required.
The weighted $3b$ events both emulate the $4b$ CR distribution and have OT couplings defined; they can thus be mapped to the SR directly.

\textbf{Section~\ref{sec:ot_unnormalized}} describes how the overall background normalization is determined.
For technical reasons, the coupling maps fit the normalized distributions, i.e., they only predict the background shapes.
The overall normalization is determined using the standard ABCD method.

 \section{The FvT Classifier for Collider Events} 
\label{app:classifier_description}
The aim of this appendix is to describe the architecture of the FvT classifier 
defined in Section~\ref{sec:classifier_method}.
Recall that our aim to design a classifier $\hat \psi$ over $\calG$, which 
\begin{enumerate}
\item[(a)] is invariant to the ordering of the constituent jets in an input event $g$;
\item[(b)] is invariant with respect to the equivalence relation $\simeq$ defined in \eqref{eq:equivalence};
\item[(c)] incorporates the dijet substructure of an event $g = g^1+g^2$. 
\end{enumerate}
\begin{figure}[!t]
\includegraphics[width=\textwidth]{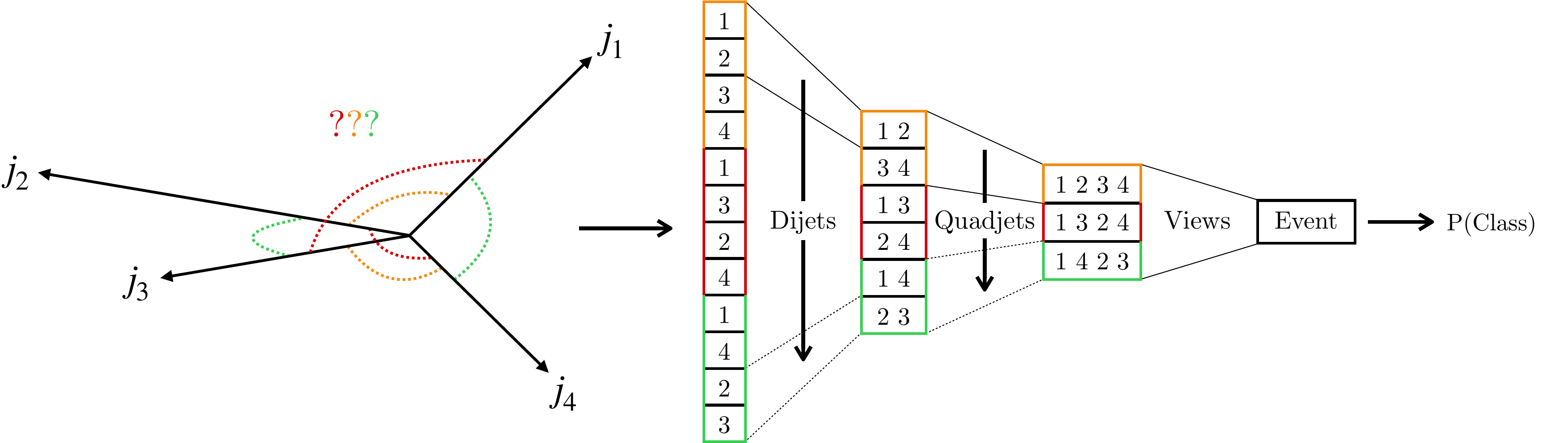}
\caption{\label{fig:resnet}
Illustration of the FvT classifier architecture. 
The classifier takes as input an event with four jets,  passes it through
the layers (1--4) of the network described below, and outputs the fitted probability that the event arose from the
$4b$ distribution as opposed to the $3b$ distribution.
}
\end{figure}
We opt for a convolutional neural network architecture with residual layers, or ResNet \citep{he2015},
as depicted in Figure \ref{fig:resnet}.
One dimensional convolutions are used to project pairs of jets into dijets and pairs of dijets into quadjets.
These convolutions are just linear maps 
\begin{equation}
  \label{eqn:convolution}
  \sum_{i,\alpha}  v_{i}^{\alpha}w_{i}^{\alpha,\beta}+b^{\beta} = v^{\beta}
\end{equation}
where $i$ is summed over the pair of input objects and $\alpha$ is summed over the input object features to produce an output in feature space indexed by $\beta$.
The learnable parameters are the $w_{i}^{\alpha,\beta}$ weight matrix and the $b^{\beta}$ bias vector. 
There is only one such set of learnable parameters for a given convolutional layer and it is applied to each pair of input objects to produce the output objects.
There are thus $I\times A\times B+B$ learnable parameters for each convolution layer 
where $I$ is the number of input objects, $A$ is the dimension of the input object 
feature space and $B$ is the dimension of the output object feature space.
An input event $g = \sum_{j=1}^4 p_{T_j} \delta_{(\eta_j,\phi_j,m_j)}$ 
to the network is treated as a one-dimensional image with 4 pixels, 
where each pixel corresponds to one jet parametrized by its coordinates
$(p_{T_j}, \eta_j, \phi_j, m_j)$. The layers of the network are then designed as follows: 
\begin{enumerate} 
\item[(i)] The 4 input pixels are duplicated twice to obtain an image with 12 pixels,
such that all 6 possible pairs of dijets appear exactly once, consecutively
in the image. Each such pair is then convolved into a one-dimensional image of six dijet pixels.
\item[(ii)] Each pair of dijets from this image is further convolved into an image with three pixels, 
consisting of the three possible pairing representations of the original four-jet, or quadjet, event.
\item[(iii)] The three quadjet pixels are combined into the last pixel representing the whole event. 
\item[(iv)] A final output layer passes the event through a
softmax activation function to produce the fitted probability $\hat \psi(g)$.
\end{enumerate}

Notice that (c) is satisfied by the construction of the network. 
We partially impose (a) by replacing pairs of dijet pixels $\{d_i,d_j\}$ in step (ii) by their sums and absolute values of their differences as follows:
\begin{equation}
\{ d_i ,\, d_j \} \rightarrow \{ (d_i+d_j)/2 ,\, |d_i-d_j|/2 \}
\end{equation}
In this way, the dijet to quadjet feature learning is invariant under the permutation
$$\{ d_i ,\, d_j \} \rightarrow \{ d_j ,\, d_i \}.$$
Similar modifications may be made in (i) to 
render the network entirely invariant to relabeling of the input jets, as in (a), but we have observed
increased performance by allowing the convolutions in step (i) to have information about the $p_T$ ordering of the jets.
Quadjet pixels $q_i^{\alpha}$ from (ii) are added together to produce a single event level pixel $e^{\alpha}$ weighted by a real-valued score $s_i$ in step (iii) to guarantee
permutation invariance of the three pixels:
\begin{equation*}
  e^{\alpha} = \sum_i s_ie_i^{\alpha}.
\end{equation*}
The quadjet scores are the softmax over quadjets of the dot product between quadjet features and a learned reference vector $w^\alpha$:
\begin{equation*}
  s_i = \text{Softmax}\left[ \sum_{\alpha} q_i^{\alpha}w^{\alpha} \right].
\end{equation*}

To enforce (b), recall that the network should be invariant 
under transformations $F_1^- : \eta \mapsto -\eta$, $F_2^-:\phi \mapsto -\phi$ and $F_2^\Delta:\phi \mapsto \phi + \Delta$, for any $\Delta \in [0,2\pi)$,
when applied simultaneously across all constituent jets of an event.
Invariance under $\eta$ flips is enforced by computing the first several layers twice, with and without the flip, and then averaging the resulting quadjet pixels prior to step (iii). 
In principle, the same may be done to enforce invariance under the rotation $F^\Delta$ by averaging
over a large grid of candidate values $\Delta$. To reduce the computational burden which
would arise from such an operation, we instead apply random rotations $\Delta$
at each batch used in the Adam optimizer \citep{kingma2017} which we choose for training the network.

In steps (i)-(iii), we also add engineered features specifically designed for dijet, quadjet, 
and event-level pixels, respectively. These engineered features are designed 
such that they preserve invariance properties (a) and (b). 
For instance, in step (i) we insert pixels containing dijet masses and 
the Euclidean distance between dijet angular variables $\eta_i, \phi_i$.

\section{Metric Approximation}
\label{app:emd}
\begin{figure}[H] 
  \includegraphics[width=0.495\textwidth]{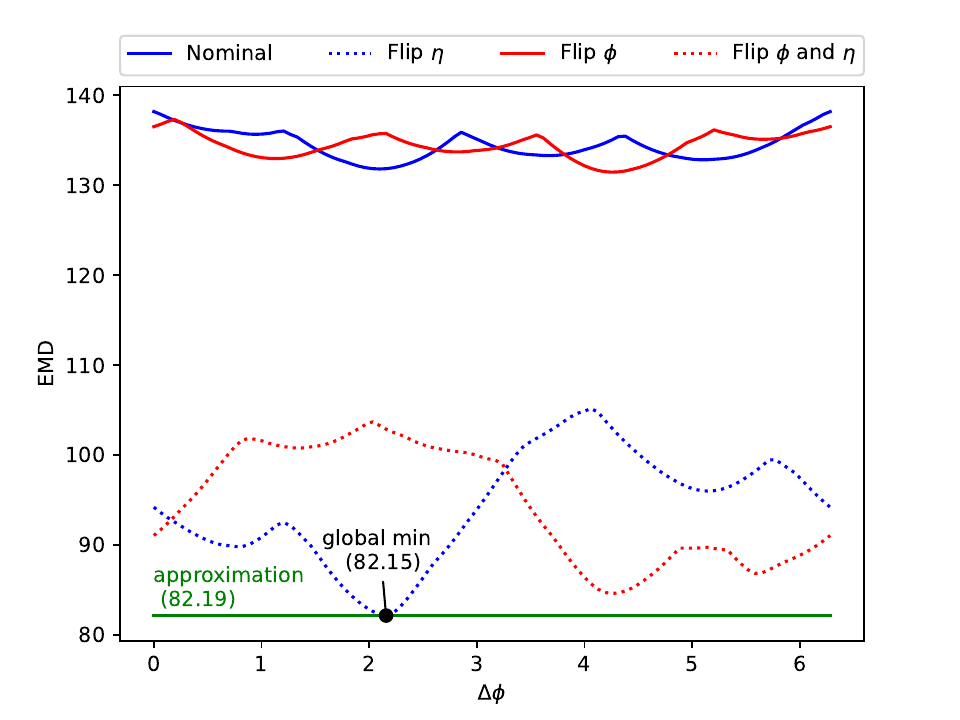}
  \includegraphics[width=0.495\textwidth]{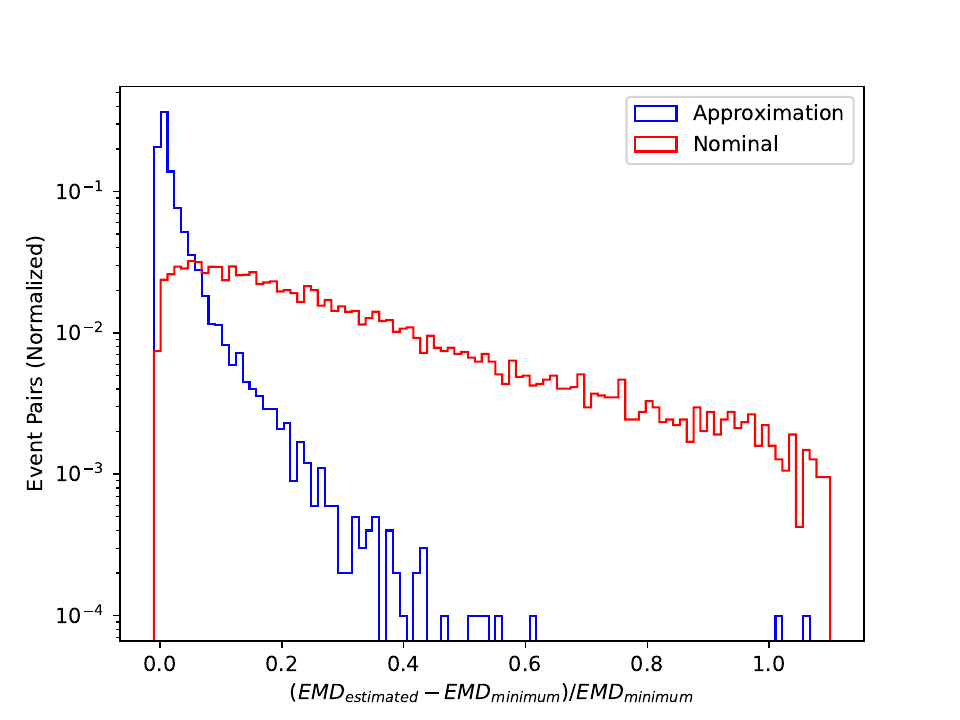}
  \vspace{-0.3in}
  \caption{ Example of the calculation of $\widetilde W$ for a single pair of events (left),
    and performance of our approximation on a sample of events (right).
    The relative bias  of our approximation is shown in blue.
    The bias using the nominal input events is shown in red.
  \label{fig:fullEMDCalc}                                                                                                                                                                                                                                                                                                                                                                                                                                                                                                                                                                                                                                                                                                                                                                                                                                                                                                                                                                                                                                                                                                                                                                                                  } 
\end{figure}

This section describes how we numerically approximate equation~\eqref{eq:final_W} taking into account the equivalence relation~$\sim$ on $\calG$.
As discussed in Section~\ref{sec:colliderEvents},  events are deemed equivalent up to relative rotations in $\phi$ and relative reflections about the $x$- and $z$- axes.
The distance between a source and a target event is taken to be the minimum value of the metric $W$ between the target event and all events in the equivalence
 class of the source event.
The left plot of Figure~\ref{fig:fullEMDCalc} shows an example of this calculation.
The metric $W$ between two events is shown as a function of the overall $\phi$ offset ($\Delta \phi$) of the source event.
The solid blue line shows the result with the nominal orientation of the axes in the source event; the dashed blue line is the result after reflecting the source event about the $z$-axis; the red curves show the results after reflection about the $x$-axis (solid red) and after a reflection of both the $x$- and $z$-axes (dashed red). 
The black dot shows the global minimum; this is the value of the final metric $\widetilde W$ defined in equation~\eqref{eq:final_W}.

A brute force calculation of the minimizer in the definiton of $\widetilde W$ is slow.
The results shown in the left plot of Figure~\ref{fig:fullEMDCalc} require 400 individual calculations of the metric $W$: 100 for each scan in $\Delta \phi$.
Each of these individual calculations is costly as they involve solving an optimal transport problem between collider events.

We speed up the calculation of the metric by approximating the value of $\widetilde W$ using an estimate of the relative $z$-axis orientation and of the $\phi$-offset of the source event that achieves the minimum in equation~\eqref{eq:final_W}. 
The orientation of the $z$-axis of the global minimum can be estimated by calculating the sum of the $z$-components of the jet momenta for each event $g = \sum_{i=1}^K p_{T_i} \delta_{(\eta_i,\phi_i,m_i)}$:
\begin{linenomath}\begin{equation}
\label{eq:pz}
S_z \equiv \sum_{i=1}^K p_{T_i}  \cos(\theta_i) = \sum_{i=1}^K p_{T_i}  \sinh(\eta_i).
\end{equation}\end{linenomath}
The global EMD minimum tends to occur when the relative orientations of the source and target event $z$-axes produce the same sign of $S_z$.
We thus reflect the $z$-axis of the source event if the $S_z$ sums of the source and target events have opposite signs.

We estimate the relative $\phi$-offset of the global minimum using the transverse thrust axis.
The transverse thrust axis is defined as the unit vector in the $(x,y)$-plane with angle $\phi_{\rm{thrust}}$ such that the sum of the jet $p_T$ projections along the unit vector are maximized:
\begin{linenomath}\begin{equation}
\label{eq:thrust}
\phi_{\rm{thrust}} = \argmax_{\phi \in [0,2\pi)} \sum_{i=1}^K p_{T_i}  \cos(\phi_i - \phi). 
\end{equation}\end{linenomath}
The relative $\phi$ rotation that achieves the minimum in the definition of $\widetilde W$ will tend to either align or anti-align the thrust axes of the events.
We estimate the value of $\widetilde W$ by taking the minimum of the $W$ calculated four times:
with the source and target thrust axes either aligned or anti-aligned, and with the source $x$-axis either nominal or reflected.
The green line in the left plot of Figure~\ref{fig:fullEMDCalc} shows the estimated value of the minimum in $\widetilde W$ using this algorithm.
Our estimation of $\widetilde W$---requiring only 4  individual evaluations of $W$---agrees well with the brute force calculation---using 400 individual EMD evaluations.

The right plot of Figure~\ref{fig:fullEMDCalc} shows the performance of our approximation on a sample of events.
The blue distribution compares the relative bias in our estimate with respect to the minimum calculated by brute force.
The majority of events have a bias of at most a few percent.
For reference, the red distribution shows the relative bias of the EMD calculated using the nominal event.



\section{Computation of Optimal Transport Couplings}
\label{app:ot_computation}

In this section, we describe our numerical approximation 
of the empirical optimal transport coupling $\hat q$ in equation~\eqref{eq:discrete_kantorovich}. 
Equation~\eqref{eq:discrete_kantorovich} is a linear program which can 
be computed exactly using simplex algorithms~\citep{peyre2019}. Such approaches have  memory complexity
which grows quadratically in $\ntc\wedge \nts$, since they require the cost matrix
$$C = (W(H_i^c, H_j^s): 1 \leq i \leq \ntc, 1 \leq j \leq \nts)$$
to be stored in memory. 
As described in Section~\ref{sec:simulation},
the sample sizes $\ntc$ and $\nts$ are at least of the order $10^5$ in our
problem, in which case the storage of the matrix $C$ becomes intractable. 
For low-dimensional problems, the storage of  $C$ can be avoided
by using the so-called back-and-forth algorithm of~\cite{jacobs2020}, which has linear
memory complexity. For higher dimensional problems, 
a common approach is to divide the datasets into
several (say, $B$) batches, and to compute $B$ separate optimal transport couplings. 
Such batches can either be obtained through subsampling or deterministic 
schemes---see~\cite{sommerfeld2019}, \cite{fatras2021c}, 
\cite{nguyen2022}, and references therein. 

We follow a similar approach in our work.  We partition the two samples
$$\calD^c = \{H_1^c, \dots, H_{\ntc}^c\}, \quad \calD^s = \{H_1^s, \dots, H_{\nts}^s\}$$
into $B \geq 1$ disjoint batches $\calD_1^c, \dots, \calD_{B}^c$ and
$\calD_1^s, \dots, \calD_{B}^s$, satisfying
$\calD^c = \bigcup_k \calD_k^c$ and $\calD^s = \bigcup_k \calD_k^s$. 
Assume for simplicity that for some $n_B^c, n_B^s \geq 1$, 
$|\calD_k^c|= \ntc^B$ and $|\calD_k^s|=\nts^B $ for all
$k=1,\dots,B$. 
We then compute the   optimal transport couplings
$$\hat q^{k} = \argmin_{(q_{ij}) \subseteq \bbR^{\ntc^B\times\nts^B}}
\sum_{H^c \in \calD_k^c} \sum_{H^s \in \calD_k^s} q_{ij} W(H^c, H^s)$$
  where the minimum is taken over all $q_{ij} \geq 0$ satisfying
  $$\sum_{i=1}^{\ntc^B} q_{ij} = \frac 1 {\nts^B}, \quad \sum_{j=1}^{\nts^B} q_{ij} 
   = \frac 1 {\ntc^B},$$
   for all $k=1, \dots, B$. We then approximate the empirical optimal transport coupling
   $\hat q$ in equation~\eqref{eq:discrete_kantorovich} by the matrix
   $$\tilde q = \frac 1 B \begin{pmatrix}
   \hat q^1  &&&\\ & \hat q^2 && \\ &&\ddots & \\ &&& \hat q^B
   \end{pmatrix}.$$
  The memory complexity of this algorithm is $O(B\ntc^B \nts^B)$
  as opposed to the complexity $O(\ntc \nts)$ incurred by any
  method which requires the storage of the matrix $C$. 
  
  In our simulations, we computed the couplings $\hat q_k$ using
  the network simplex solver described by~\cite{bonneel2011}, as implemented
  in the Python Optimal Transport package~\citep{flamary2021}. 
  Our simulations were run on a standard Linux machine with 12 cores
  and 32GB of RAM.   
  We chose $B=16$, which is approximately the smallest
  value of $B$ for which the computation of the couplings $\hat q^k$
  did not exceed our machine's memory limit.  
  
  We chose the batches according to the following procedure. For any
  event $G = \sum_{j=1}^4 p_{T_j} \delta_{(\eta_j,\phi_j,m_j)}$, 
  let $s_T(G) = \sum_{j=1}^4 p_{T_j}$ denote the scalar sum of the transverse
  momenta of $G$. Let $H_{(i)}^c$ denote the event among $H_1^c, \dots, H_{\ntc}^c$
  with the $i$-th smallest $s_T$ value, for all $i=1, \dots, \ntc$. That is, 
  $$s_T(H_{(1)}^c) \leq s_T(H_{(2)}^c) \leq \dots \leq s_T(H_{(\ntc)}^c).$$
  We likewise define $H_{(j)}^s$ for $j=1,\dots,\nts$ such that 
  $$s_T(H_{(1)}^s) \leq s_T(H_{(2)}^s) \leq \dots \leq s_T(H_{(\nts)}^s).$$  
  We then set, for $k=1, \dots, B$, 
  $$\calD_k^c = \{H_{(B (r -1) + k)}^c : 1 \leq r \leq \ntc^B\},\quad 
    \calD_k^s = \{H_{(B (r -1) + k)}^s : 1 \leq r \leq \nts^B\}.$$
  This choice ensures that each batch contains events with
  a comparable range of $s_T$ values. We impose this property 
  because the penalty  term in the definition of $\widetilde W$ (Eq.~\eqref{eq:emd}) 
  is sensitive to large deviations of $s_T$ values.
  We leave open for future work
  whether different batching methods, such as subsampling, would yield
  improved performance.

\clearpage 

\section{Additional Simulations Results}
\label{app:additional_simulation_results}
In this appendix, we report additional results from the simulation study in Section~\ref{sec:simulation}.

\begin{figure}[H] 
\includegraphics[width=0.495\textwidth]{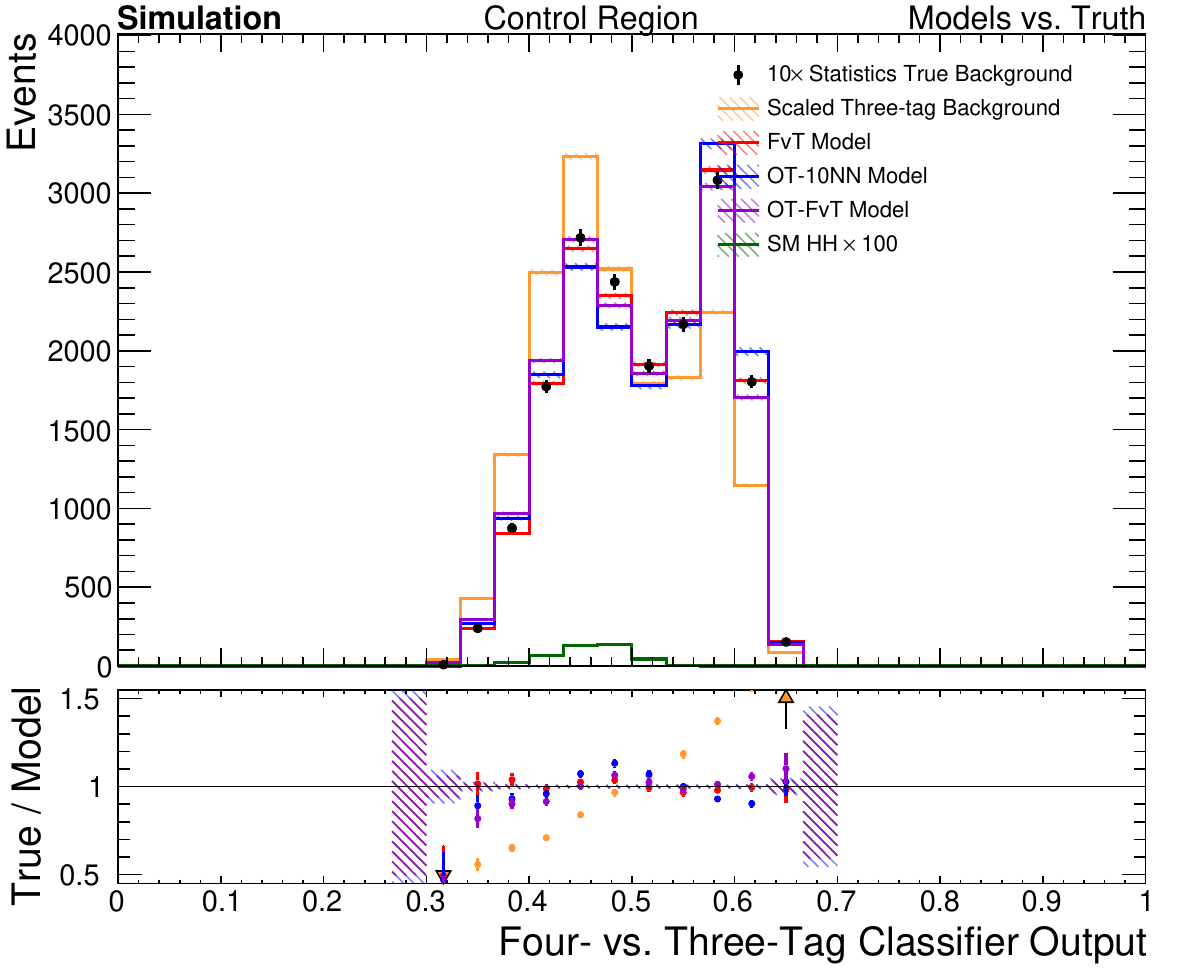}
\includegraphics[width=0.495\textwidth]{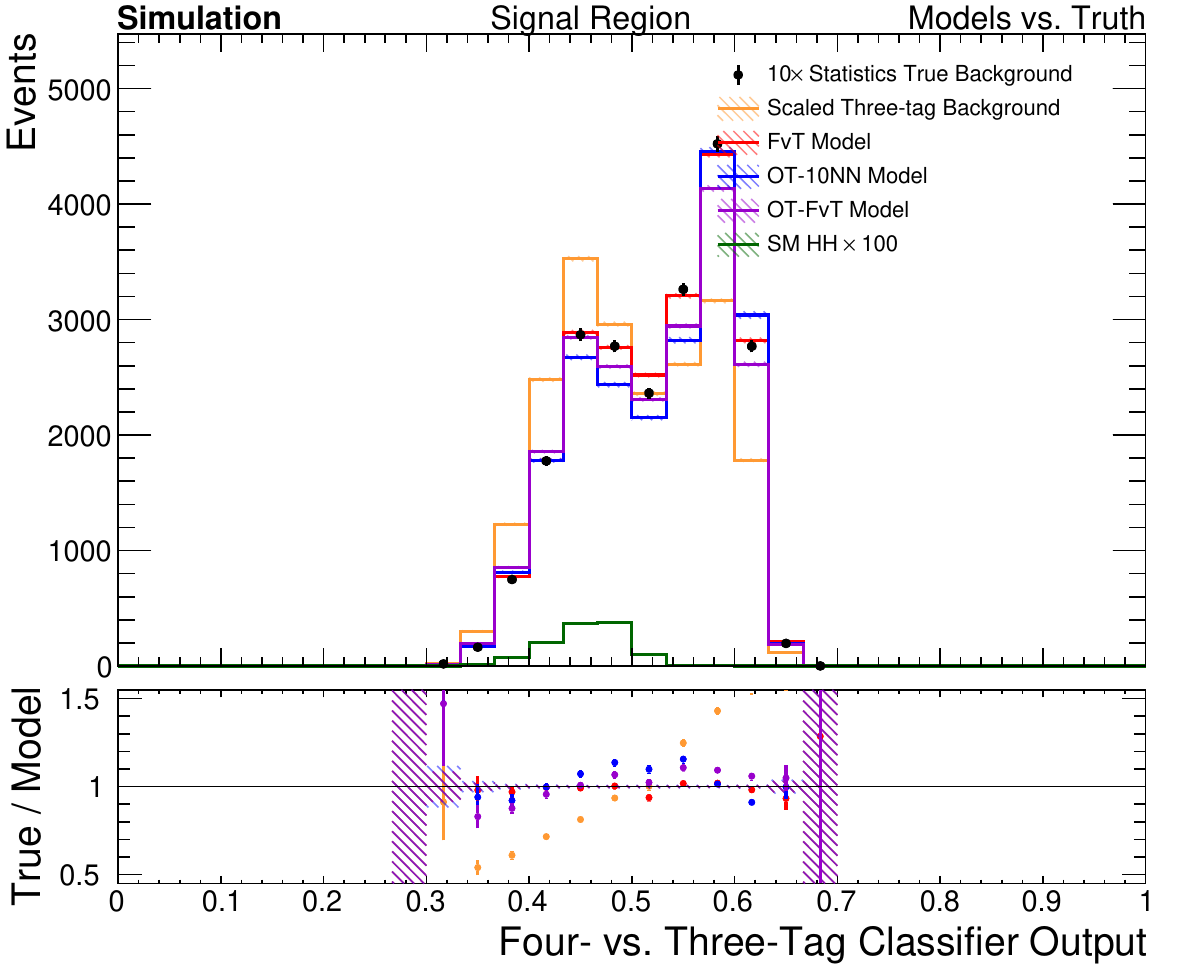}
\caption{ Histograms of the FvT classifier output variable in the Control Region (left) and the Signal
Region (right), for
the three background models as well as the upsampled 4b data (treated as the ground truth), 
the 3b data (normalized by the factor $\nts\nfc/\ntc$), and 
the di-Higgs signal sample (SM HH).  \label{fig:sim_FvT_log}.} 
\end{figure}

\begin{figure}[H] 
\includegraphics[width=0.495\textwidth]{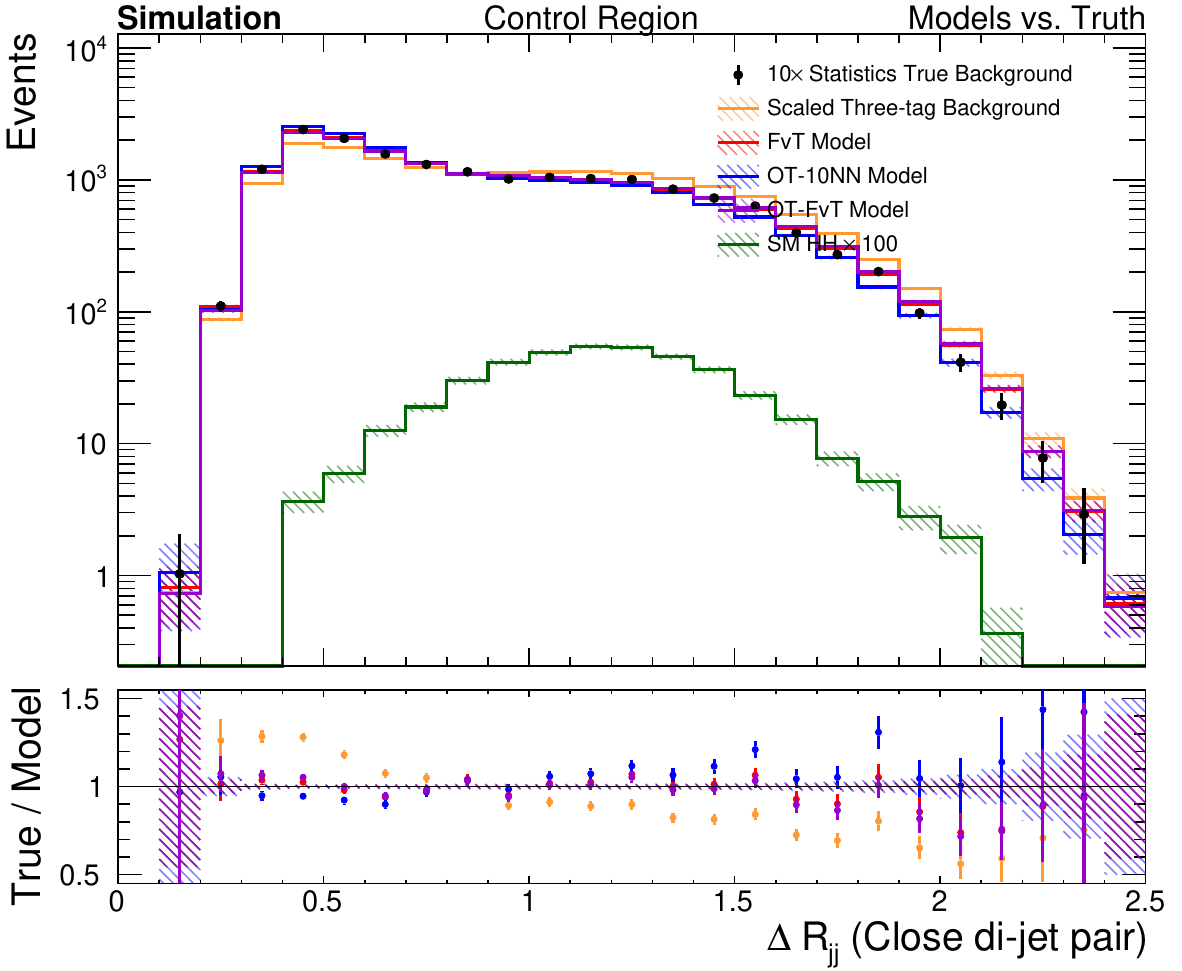}
\includegraphics[width=0.495\textwidth]{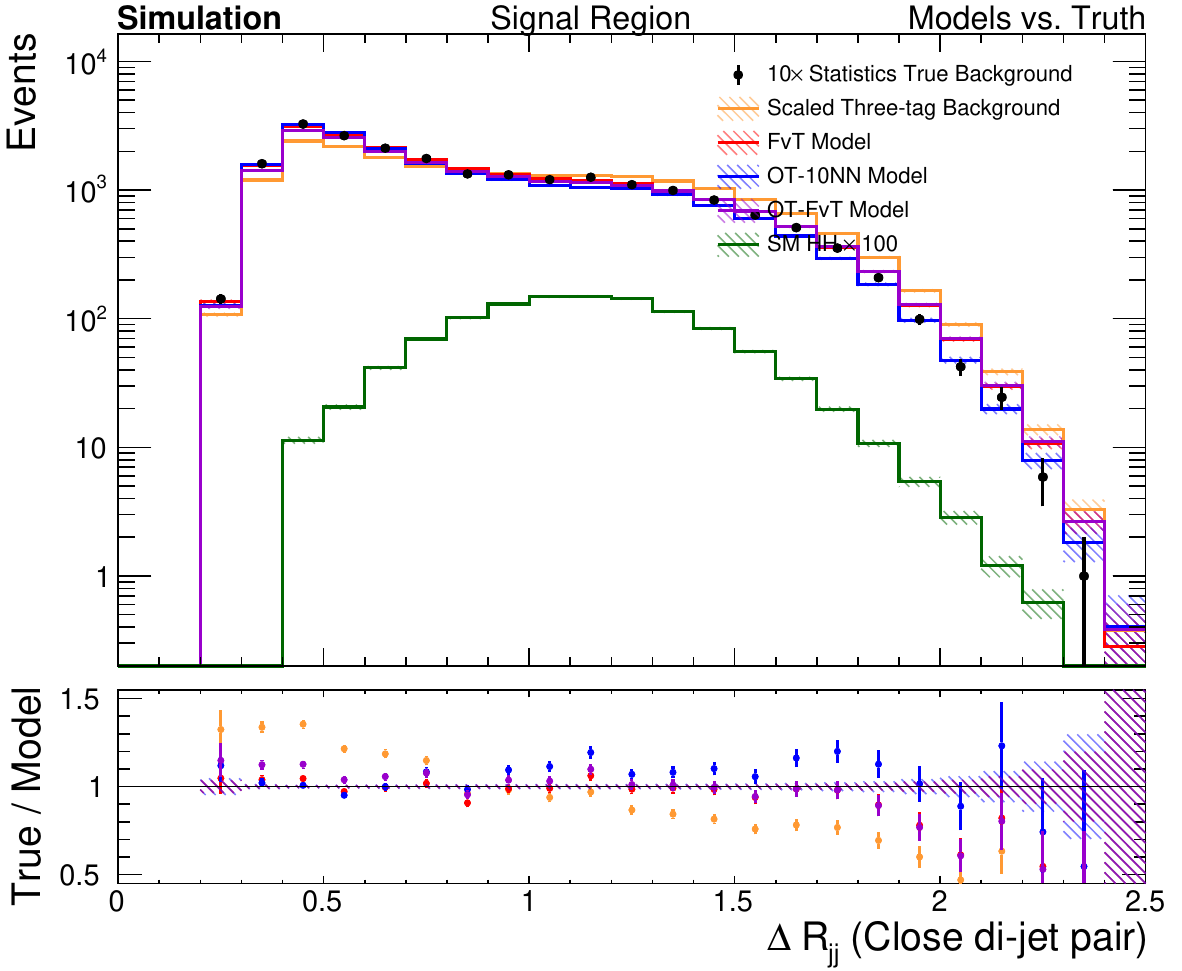}
\caption{ Histograms of the ``$\Delta R_{jj}$--Close'' variable
(defined as the angular
distance between the two closest jets of an event)
in the Control Region (left) and the Signal
Region (right), for
the three background models as well as the upsampled 4b data (treated as the ground truth), 
the 3b data (normalized by the factor $\nts\nfc/\ntc$), and 
the di-Higgs signal sample (SM HH).  \label{fig:sim_mHH_dRjjOther_log}.} 
\end{figure}

\begin{figure}[H] 
\includegraphics[width=0.495\textwidth]{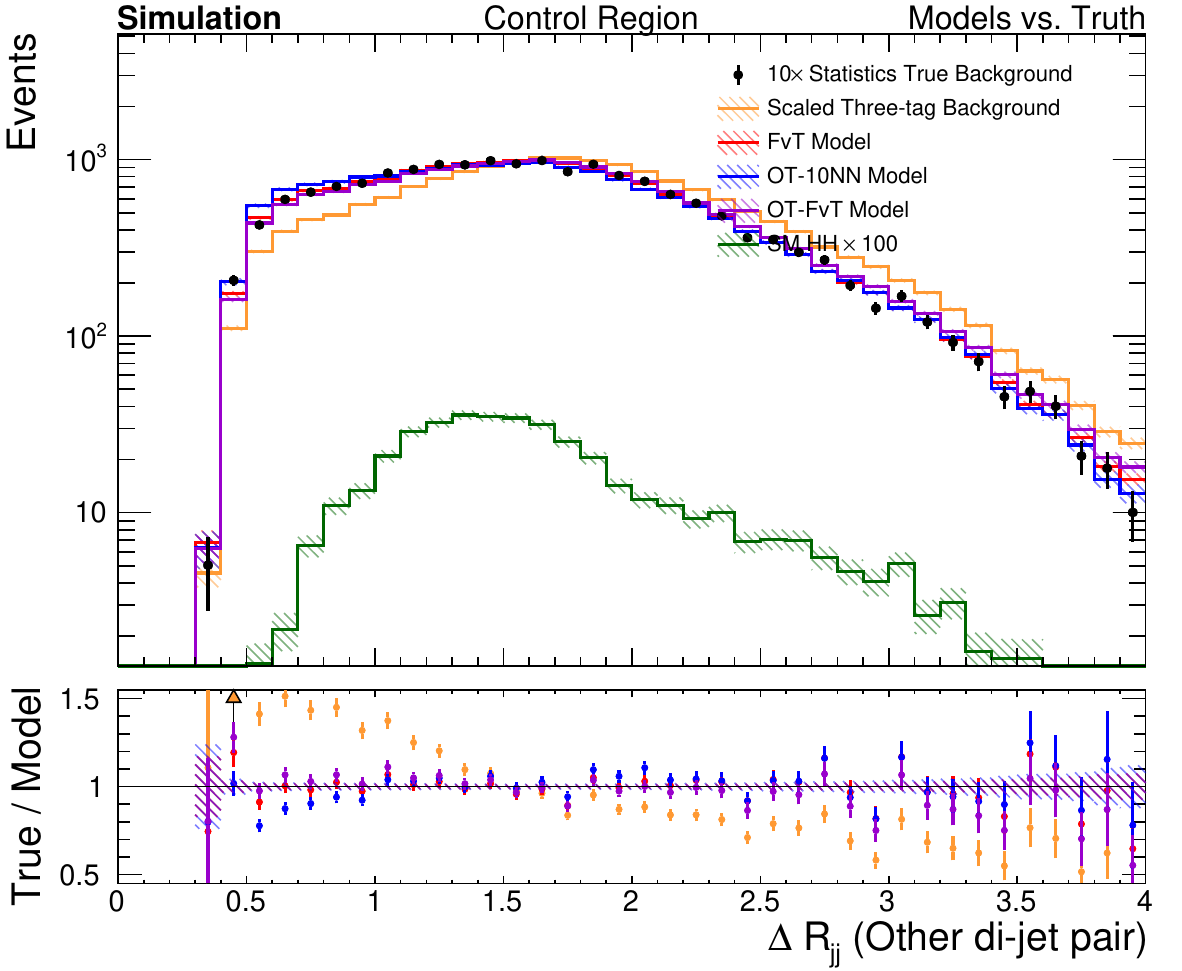}
\includegraphics[width=0.495\textwidth]{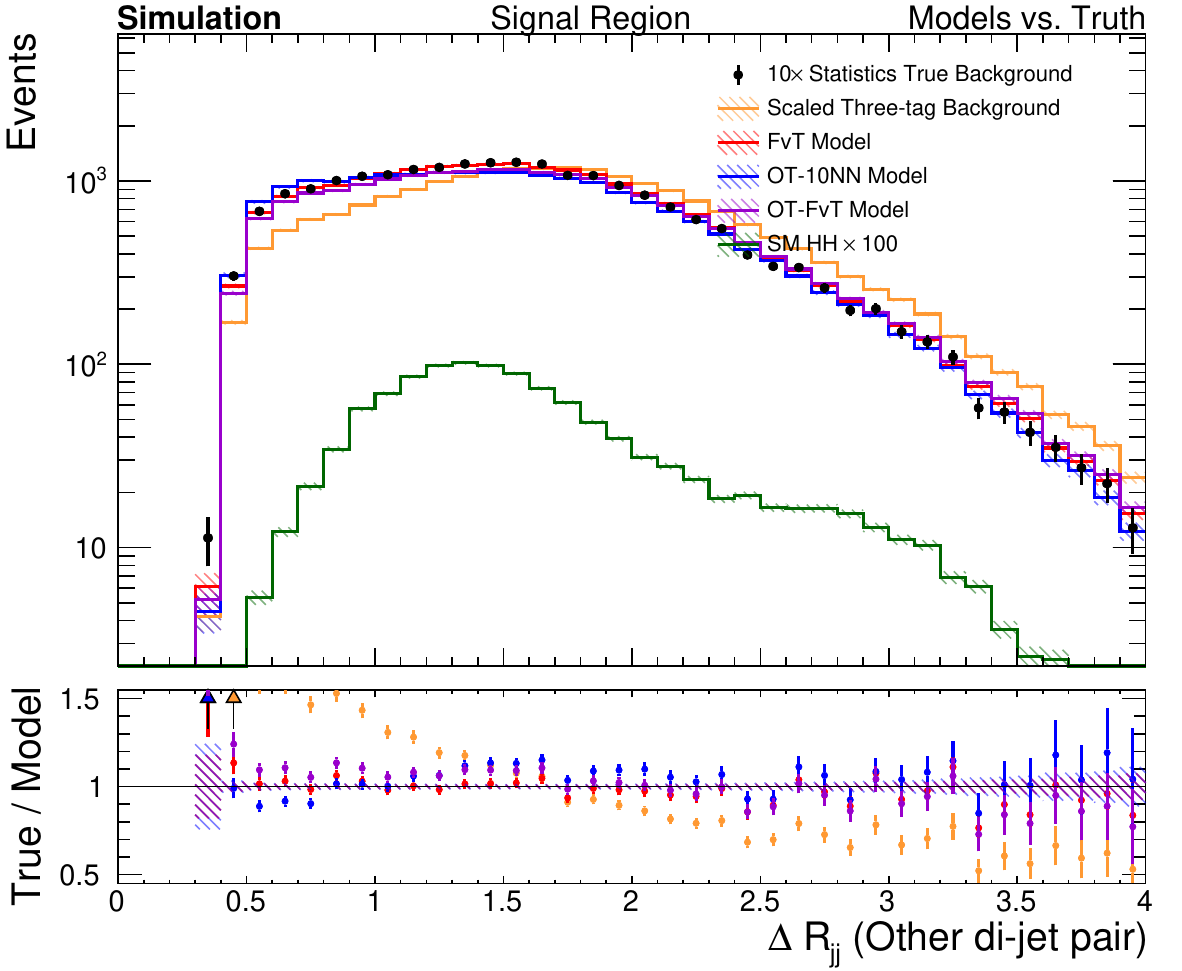}
\caption{ Histograms of the ``$\Delta R_{jj}$--Other'' variable (defined
as the angular distance between the two jets of an event other than the two which are closest) in the Control Region (left) and the Signal
Region (right), for
the three background models as well as the upsampled 4b data (treated as the ground truth), 
the 3b data (normalized by the factor $\nts\nfc/\ntc$), and 
the di-Higgs signal sample (SM HH).  \label{fig:sim_mHH_dRjjOther_log}.} 
\end{figure}

 \bibliographystyle{imsart-nameyear} 
\bibliography{manuscript}
 
\end{document}